\documentclass [12pt]{article}
\usepackage {graphicx}
\usepackage {amssymb}
\usepackage {amsmath}
\usepackage {longtable}

\title{Nonlinear Dynamo}
\author{$^{1,2}$\textbf{M.I.Kopp}, $^{3}$\textbf{A.V.Tur} $^{1,2}$\textbf{V.V.Yanovsky} }

\begin{document}

 \maketitle

$^{1}$\textit{Institute for Single Crystals, NAS Ukraine, Nauky Ave. 60, Kharkov 31001, Ukraine}

$^{2}$ \textit{V.N. Karazin Kharkiv National University 4 Svobody Sq., Kharkov 61022, Ukraine}

$^{3}$\textit{Universite de Toulouse [UPS], CNRS, Institut de Recherche en Astrophysique et Planetologie, 9 avenue du Colonel Roche, BP 44346, 31028 Toulouse Cedex 4, France}

\bigskip

\begin{abstract}In this manuscript using the asymptotic method of multiscale nonlinear theory we construct  a nonlinear theory of the appearance of large-scale structures in the stratified conductive medium with the presence of small-scale oscillations of the velocity field  and magnetic fields. These small-scale stationary oscillations are maintained by small external sources at low Reynolds numbers. We obtain a nonlinear system of equations describing the evolution of large-scale structures of the velocity field  and magnetic fields. The linear stage of evolution leads to the known instability. In this article we consider the stationary large-scale structures of a magnetic field arising at stabilization of linear instability.
\end{abstract}

\section{Introduction}
\label{on}

It is known, that the term “dynamo” originated in theory of the generation of large-scale magnetic fields in magnetic hydrodynamics, mainly in its applications to astrophysics and geophysics. The main mechanism of the generation of large-scale magnetic fields (i.e., the dynamo mechanism) is a violation of the parity of fluid velocity field.  In particular, a violation of parity leads to a non-zero velocity field helicity $\alpha = < \vec{V} rot \vec{V}>$ ($\alpha$-effect). Here  α is pseudo scalar . More generally, the violation of parity gives AKA effect (anisotropic kinetic α effect). In turn, these effects generate large-scale magnetic field instability and generate large-scale vortex instability of fluid. In recent years, this effect got the name of vortex dynamo.  Currently the theory of magnetic dynamo is well developed in the kinematic formulation, when we assume that the velocity field is given. This theory is presented in numerous articles and books (e.g. \cite{s21}, \cite{s24}, \cite{s26}, \cite{s27}, \cite{s28}, \cite{s35}). The theory of the vortex dynamo is younger, but it is also the subject of many papers (e.g. \cite{s10}-\cite{s13}). The main application of the vortex dynamo theory is the generation of localized vortex structures in the atmosphere, for example, such as tropical cyclones \cite{s11}, \cite{s13}. Actually the large-scale vortex structures can be generated by the same dynamo mechanism in the conductive fluid, like the magnetic field. The question then, is how these vortex structures coexist with the magnetic structures and what is the nonlinear dynamics of magnetic and vortex structures.  These issues are the subject of this article.

In this paper, we consider that a small-scale external force with a nonzero helicity operates in the conductive fluid, which engenders the helical velocity field fluctuations, i.e. models helical turbulence.
Note that we understand under the large scale structure the structures which the characteristic scale is much larger than the scale of external force or of turbulence which generates them.
As a mathematical technique which allows to study the nonlinear dynamo we use the method of multiscale developments proposed in \cite{s15}. In this method, the asymptotic development is carried out according to the small Reynolds number $R \ll 1$. This makes possible to obtain the closed equations for large-scale fields and study the linear and nonlinear instabilities. As a result, we observe a large number of different nonlinear vortices and magnetic structures: nonlinear waves, solitons, kinks and chaotic structures.

Strictly speaking, this theory is applicable at small Reynolds numbers, i.e. for motion of conducting fluid with a sufficiently high viscosity, for example a liquid core of the Earth. Actually in astrophysics, the Reynolds number is usually large and the motion of conducting fluid is turbulent. If we use the turbulent viscosity to estimate the Reynolds number, the effective Reynolds number can be reduced to a value of the order of unity. It enables to use the theory presented in this article for qualitative estimations. Some examples of such assessments to the photosphere of the Sun will be given further.

The paper is organized as follows. In Sec.\ref{to} are set out the basic equations and the formulation of the problem. In Sec.\ref{th} are derived the basic equations of the nonlinear dynamo. In Sec.\ref{fo} is described the large-scale linear instability. In Sec.\ref{fi} are described the stationary nonlinear magnetic structures.  In the appendices are described various technical issues, such as multi-scale asymptotic developments and the closure of the Reynolds stresses.

\section{Basic equations and formulation of the problem}
\label{to}

Let us consider equations of motion of the incompressible electroconducting medium with a constant temperature gradient in Boussinesq approximation \cite{s7}:
\begin{equation}\label{e1}
  \frac{{\partial \vec V}}{{\partial t}} + \left( {\vec V\nabla } \right)\vec V =  - \frac{{\nabla {\rm{P}}}}{{{\rho _{00}}}} + \nu \Delta \vec V + \frac{1}{{4\pi {\rho _{00}}}}\left[ {rot\vec B \times \vec B} \right] + g\beta T\vec e + {\vec F_0},
\end{equation}
\begin{equation}\label{e2}
  \frac{{\partial T}}{{\partial t}} + \left( {\vec V\nabla } \right)T = \chi \Delta T - {V_z}A,
\end{equation}
where $\vec e = (0,0,1)$ is the unit vector in the direction of the axis $z$, $\beta$ is the thermal development coefficient, $A = \frac{{d{T_{00}}}}{{dz}}$ is the constant temperature gradient, $A = const$, $A > 0$. To the equations (\ref{e1})-(\ref{e2}) should be added the equations for magnetic induction $\vec B$ and solenoidality conditions for fields $\vec V$ and $\vec B$:
\begin{equation}\label{e3}
  \frac{{\partial \vec B}}{{\partial t}} = rot\left[ {\vec V \times \vec B} \right] + {\nu _m}\Delta \vec B\;
\end{equation}
\begin{equation}\label{e4}
  div\vec B = 0,\quad \quad div\vec V = 0
\end{equation}
Here, ${\nu _m} = \frac{c^2}{{4\pi \sigma }}$ is the the coefficient of magnetic viscosity, $\sigma$ is the electrical conductivity of the medium. The system of equations (\ref{e1})-(\ref{e4}) describes the evolution of disturbances on the background of the equilibrium state defined by the equilibrium condition:
\begin{equation}\label{e5}
  \nabla {P_{00}} = {\rho _{00}}g\beta {T_{00}}
\end{equation}
${\rho _{00}} = const$ is the density of medium, $\chi$ is the thermal conductivity of the medium. We chose these unusual designations for an equilibrium state to  avoid further any  confusion with asymptotic development. Equation (\ref{e1}) contains  the external force ${\vec F_0}$ which has the following property:
\begin{equation}\label{e6}
  {\vec F_0} = {f_0}{\vec F_0}\left( {\frac{x}{{{\lambda _0}}};\frac{t}{{{t_0}}}} \right),\quad \quad div{\vec F_0} = 0,\quad \quad {\vec F_0} \cdot rot{\vec F_0} \ne 0
\end{equation}
where ${\lambda _0}$ - the characteristic scale, ${t_0}$ - the characteristic time, ${f_0}$ - the characteristic amplitude of the force. The main role of this force is to create the helical velocity field ${\vec v_0}$ with a low Reynolds number $R = \frac{{{v_0}{t_0}}}{{{\lambda _0}}} \ll 1$ in  the medium of small-scale fluctuations. In other words, to maintain the small-scale  helical turbulence. It is easy to notice that the characteristic velocity ${v_0}$, generated  by an external force, has the same characteristic scales:
\begin{equation}\label{e7}
  {v_0} = {v_0}\left( {\frac{x}{{{\lambda _0}}},\frac{t}{{{t_0}}}} \right)
\end{equation}
Now we write the system of equations (\ref{e1})-(\ref{e4}) in dimensionless variables:
\begin{equation}\label{e8}
  \vec x \to \frac{{\vec x}}{{{\lambda _0}}},\quad t \to \frac{t}{{{t_0}}}, \quad {\vec v_0} \to \frac{{{{\vec v}_0}}}{{{v_0}}}, \quad {\vec F_0} \to \frac{{{{\vec F}_0}}}{{{f_0}}}, \quad P \to \frac{P}{{{\rho _{00}}{P_0}}},
\end{equation}
\[\vec B \to \frac{{\vec B}}{{{B_0}}}, \quad {t_0} = \frac{{\lambda _0^2}}{\nu }, \quad {P_0} = \frac{{{v_0}\nu }}{{\lambda _0^2}}\;,\;\;T \to \frac{T}{{{\lambda _0}A}}\;\]
In these variables, equations (\ref{e1})-(\ref{e4}) take the form:
\begin{equation}\label{e9}
  \frac{{\partial \vec V}}{{\partial t}} + R \left( {\vec V \nabla } \right) \vec V =  - \nabla P + \Delta \vec V + \widetilde {Ra}T \vec e + \widetilde Q R \left[ {rot\vec B \times \vec B} \right] + {\vec F_0}
\end{equation}
\begin{equation}\label{e10}
  \frac{{\partial T}}{{\partial t}} - P{r^{ - 1}}\Delta T =  - R\left( {\vec V\nabla } \right)T - {V_z}
\end{equation}
\begin{equation}\label{e11}
  \frac{{\partial \vec B}}{{\partial t}} - P{m^{ - 1}}\Delta \vec B = R{\kern 1pt} rot\left[ {\vec V \times \vec B} \right]
\end{equation}
where $\widetilde {Ra} = \frac{{Ra}}{{Pr}}$, $Ra = \frac{{g\beta A\lambda _0^4}}{{\nu \chi }}$ is the Rayleigh number on the scale of ${\lambda _0}$; $Pr = \frac{\nu }{\chi }$ is the Prandtl number; $\widetilde Q = \frac{Q}{{Pm}}$, $Q = \frac{{\sigma B_0^2\lambda _0^2}}{{{c^2}{\rho _{00}}\nu }}$ is the Chandrasekhar number of scale ${\lambda _0}$; $Pm = \frac{\nu }{{{\nu _m}}}$ is the magnetic Prandtl number; ${B_0}$ - the characteristic small-scale magnetic field, which we consider as initial small magnetic field \cite{s26} ).
Let us consider as the small parameter of the asymptotic development the  Reynolds number $R$ of the small-scale turbulence. Smallness of the remaining parameters is not supposed and the parameters of $Ra$ and $Q$, do not influence the scheme of asymptotic development.

We turn now to the next formulation of the problem. Let the external force ${F_0}$ be helical, small-scale and high-frequency.  This force engenders the small-scale fluctuations of velocity and temperature compared to the equilibrium state. Small-scale fluctuations of a magnetic field are generated by non turbulent  mechanisms, for example, by means of thermo-effects \cite{s34}, \cite{s35}, plasma instabilities \cite{s36}, \cite{s37}, etc. When averaging the rapidly oscillating small-scale fluctuations give zero. However, due to the nonlinear interaction between them, there may be terms which do not vanish in the averaging. These terms are called secular, and they are condition of solvability of multi-scale asymptotic developpement. So the main problem is to find and study the solvability equations, i.e., the equations for large-scale perturbations.

\section{Equations of non-linear dynamo in "quasi two-dimensional" model}
\label{th}

Let us consider in more detail the application of the method of multi-scale asymptotic development to the problem of nonlinear evolution of the large-scale vortex and magnetic perturbations in the convective and electrically conductive medium. In order to  construct the multi-scale asymptotic developments we will use the methods of \cite{s18}, \cite{s19}. Let us denote the small-scale variables ${x_0} = ({\vec x_0},{t_0})$, and the large-scale $X = (\vec X,T)$. The derivative of $\frac{\partial }{{\partial x_0^i}}$ is denoted by ${\partial _i}$, and the derivative $\frac{\partial }{{\partial {t_0}}}$ as ${\partial _t}$. Further, the large-scale spatial and temporal derivatives will be denoted as:
\[\frac{\partial }{{\partial {X_i}}} \equiv {\nabla _i} ,\]
\[\frac{\partial }{{\partial T}} \equiv {\partial _T} .\]
In accordance with the method of multiple scales represent the spatial and temporal derivatives in equations (\ref{e9})-(\ref{e11}) in the form of derivatives of the small-scale and large-scale variables:
\begin{equation}\label{e12}
  \frac{\partial }{{\partial {x_i}}} \to {\partial _i} + {R^2}{\nabla _i}
\end{equation}
\begin{equation}\label{e13}
  \frac{\partial }{{\partial t}} \to {\partial _t} + {R^4}{\partial _T}
\end{equation}
Now the perturbed fields $\vec V$, $T$, $\vec B$ and $P$ we develop into series over the small parameter $R$ and we obtain:
	\[\vec V(\vec x,t)\;\; = \frac{1}{R}{W_{ - 1}}\left( X \right) + {\vec v_0}\left( {{x_0}} \right) + R{\vec v_1} + {R^2}{\vec v_2} + {R^3}{\vec v_3} +  \cdots \]
	\[T(\vec x,t) = \frac{1}{R}{T_{ - 1}}\left( X \right) + {T_0}\left( {{x_0}} \right) + R{T_1} + {R^2}{T_2} + {R^3}{T_3} +  \cdots \]
	\[\vec B\left( {\vec x,t} \right) = \frac{1}{R}{\vec B_{ - 1}}\left( X \right) + {\vec B_0}\left( {{x_0}} \right) + R{\vec B_1} + {R^2}{\vec B_2} + {R^3}{\vec B_3} +  \cdots \]
\begin{equation}\label{e14}
  P(\vec x,t) = \frac{1}{{{R^3}}}{P_{ - 3}} + \frac{1}{{{R^2}}}{P_{ - 2}} + \frac{1}{R}{P_{ - 1}} + {P_0} + R({P_1} + {\overline P _1}\left( X \right)) + {R^2}{P_2} + {R^3}{P_3} +  \cdots
\end{equation}
Here the contribution of ${P_1}$ depends on small-scale variables and ${\overline P _1}\left( X \right)$ only on large scale. Substituting (\ref{e12})-(\ref{e14}) in the system of equations (\ref{e9})-(\ref{e11}) and putting together terms of the same order for the $R$ including ${R^3}$, we obtain a system of equations of multi-scale asymptotic developments. The main problem is to isolate from these equations the secular conditions which determine the dynamics of disturbances on a large scale. The algebraic structure of the asymptotic development of the equations (\ref{e9})-(\ref{e11}) in different orders of $R$ is given in Appendix I. It is also shown that the main secular equation, i.e. equation for large-scale fields  is obtained in the order ${R^3}$:
\begin{equation}\label{e15}
  {\partial _T}{W_i} - \Delta {W_i} + {\nabla _k}\overline {(v_0^kv_0^i)}  =  - {\nabla _i}\overline P  + \widetilde Q\left( {{\nabla _k}\left( {\overline {B_0^iB_0^k} } \right) - \frac{{{\nabla _i}}}{2}{{\left( {\overline {B_0^k} } \right)}^2}} \right)
\end{equation}
\begin{equation}\label{e16}
  {\partial _T}{H_i} - P{m^{ - {1}}}\Delta {H_i} = {\nabla _j}\left( {\overline {v_0^iB_0^j} } \right) - {\nabla _j}\left( {\overline {v_0^jB_0^i} } \right)
\end{equation}
\begin{equation}\label{e17}
  {\partial _T}\Theta  - P{r^{ - 1}}\Delta \Theta  + {\nabla _k}\left( {\overline {v_0^k{T_0}} } \right) = 0
\end{equation}
Secular equations, derived in Appendix I are added to the Equations (\ref{e15})-(\ref{e17}):
\[{\nabla _k}\left( {{W_k}{W_i}} \right) =  - {\nabla _i}{\overline P _{ - 1}} + \widetilde Q({\nabla _k}{H_i} - {\nabla _i}{H}_k ){H_k}\]
\[{\nabla _k}\left( {{W_k}\Theta } \right) = 0\]
\[{W_j}{\nabla _j}{H_i}= {H_j}{\nabla _j}{W_i}\]
\[{\nabla _i}{W_i} = 0,\quad \quad {\nabla _i}{H_i} = 0,\quad \quad {W_z} = 0\]
Thus, to obtain  equations (\ref{e15})-(\ref{e16}), which describes the evolution of large-scale fields $\vec W$ and $\vec H$ it is necessary to reach the third order of perturbation theory. This is a quite typical phenomenon when applying the method of multiscale developments. Equations (\ref{e15})-(\ref{e16}) become closed after calculation of the correlation functions of Reynolds stress:
	\[\overline {v_0^k v_0^i}  = \overline {v_{01}^k{{(v_{01}^i)}^*}}  + \overline {{{(v_{01}^k)}^*}v_{01}^i}  + \overline {v_{03}^k{{(v_{03}^i)}^*}}  + \overline {{{(v_{03}^k)}^*}v_{0}3}^i  = T_{(1)}^{ki} + T_{(2)}^{ki},\]
Maxwell stresses:
	\[\overline {B_0^iB_0^k}  = \overline {B_{01}^i{{(B_{01}^k)}^*}}  + \overline {{{(B_{01}^i)}^*}B_{01}^k}  + \overline {B_{03}^i{{(B_{03}^k)}^*}}  + \overline {{{(B_{{0}3}^i)}^*}B_{03}^k}  = S_{(1)}^{ik} + S_{(2)}^{ik},\]
and correlators entering in the definition of turbulent e.m.f ${{\cal E}_n} = {\varepsilon _{nij}}\overline {v_0^iB_0^j}$ (see for example \cite{s26}):
	\[\overline {v_0^iB_0^j}  = \overline {v_{01}^i{{(B_{01}^j)}^*}}  + \overline {{{(v_{01}^i)}^*}B_{01}^j}  + \overline {v_{03}^i{{(B_{03}^j)}^*}}  + \overline {{{(v_{03}^i)}^*}B_{03}^j}  = G_{(1)}^{ij} + G_{(2)}^{ij}.\]
Complex conjugate values here and later will be designated by an asterisk. The "quasi two-dimensional" approach is often used  in many astrophysical and geophysical problems to describe the dynamics of large-scale vortex and magnetic fields \cite{s1}, \cite{s18}, \cite{s19}, \cite{s35}. In the frame of this approximation, we assume that the  large-scale derivative over $Z$ is much more than others derivatives, i.e.,
	\[\quad \quad \frac{\partial }{{\partial Z}} \gg \frac{\partial }{{\partial X}}, \,\, \frac{\partial }{{\partial Y}},\]
and the geometry of the large-scale fields is as follows:
\begin{equation}\label{e18}
  \vec W = \left( {{W_1}\left( Z \right),\;{W_2}\left( Z \right),\;0} \right), \quad \vec H = \left( {{H_1}\left( Z \right),\;{H_2}\left( Z \right),\;0} \right)
\end{equation}
This geometry corresponds to the large-scale field (Beltrami field): $\vec W \times rot\vec W = 0$ and $\vec H \times rot\vec H = 0$. Finally, taking into account the geometry of the problem (\ref{e18}), the equation for large-scale disturbances take the form:
\begin{equation}\label{e19}
  {\partial _T}{W_1} - \nabla _Z^2{W_1} + {\nabla _Z}\left( {\overline {v_0^zv_0^x} } \right) = \widetilde Q{\nabla _Z}(\overline {B_0^zB_0^x} )
\end{equation}
\begin{equation}\label{e20}
  {\partial _T}{W_2} - \nabla _Z^2{W_2} + {\nabla _Z}\left( {\overline {v_0^zv_0^y} } \right) = \widetilde Q{\nabla _Z}(\overline {B_0^zB_0^y} )
\end{equation}
\begin{equation}\label{e21}
  {\partial _T}{H_1} - P{m^{ - 1}}\nabla _Z^2{H_1} = {\nabla _Z}\left( {\overline {v_0^xB_0^z} } \right) - {\nabla _Z}(\overline {v_0^zB_0^x} )
\end{equation}
\begin{equation}\label{e22}
  {\partial _T}{H_2} - P{m^{ - 1}}\Delta {H_2} = {\nabla _Z}\left( {\overline {v_0^yB_0^z} } \right) - {\nabla _Z}(\overline {v_0^zB_0^x} )
\end{equation}
\begin{equation}\label{e23}
  {\partial _T}\Theta  - P{r^{ - 1}}\nabla _Z^2\Theta  + {\nabla _Z}\left( {\overline {v_0^z{T_0}} } \right) = 0
\end{equation}
\begin{equation}\label{e24}
  \widetilde {Ra}\Theta {e_z} = {\nabla _Z}{P_{ - 3}},\;\;{\nabla _Z} \equiv \frac{\partial }{{\partial Z}}
\end{equation}
For equations (\ref{e19})-(\ref{e22}) in closed form, we use solutions of equations for small-scale fields in zero order of $R$, obtained in Appendix II. Further it is necessary to calculate the correlators included in the system of equations (\ref{e19}) - (\ref{e22}). The technical aspect of this issue is described in detail in Annex III. in Appendix III. As a result of these the calculations of components $T_{(2)}^{31}$, $T_{(1)}^{32}$, $S_{(2)}^{31}$, $S_{(1)}^{32}$, $\delta {G_{(2)}} = G_{(2)}^{13} - G_{(2)}^{31}$, $\delta {G_{(1)}} = G_{(1)}^{23} - G_{(1)}^{32}$, we get a closed equation for large scale velocity fields $({W_1},{W_2})$ and magnetic field $({H_1},{H_2})$ in the following form:
\begin{equation}\label{e25}
  {\partial _T}{W_1} - \nabla _Z^2{W_1} - {\nabla _Z}\left[ {{\alpha ^{(2)}} \cdot \left( {1 - {W_2}} \right)\left( {1 - \frac{{H_2^2PmQ}}{{\left( {1 + P{m^2}{{\left( {1 - {W_2}} \right)}^2}} \right)}}} \right)} \right] = 0
\end{equation}
\begin{equation}\label{e26}
  {\partial _T}{W_2} - \nabla _Z^2{W_2} + {\nabla _Z}\left[ {{\alpha ^{(1)}} \cdot \left( {1 - {W_1}} \right)\left( {1 - \frac{{H_1^2PmQ}}{{\left( {1 + P{m^2}{{\left( {1 - {W_1}} \right)}^2}} \right)}}} \right)} \right] = 0
\end{equation}
\begin{equation}\label{e27}
  {\partial _T}{H_1} - P{m^{ - 1}}\nabla _Z^2{H_1} - {\nabla _Z}\left( {\alpha _H^{(2)} \cdot {H_2}} \right) = 0
\end{equation}
\begin{equation}\label{e28}
  {\partial _T}{H_2} - P{m^{ - 1}}\nabla _Z^2{H_2} + {\nabla _Z}\left( {\alpha _H^{(1)} \cdot {H_1}} \right) = 0
\end{equation}
Equations (\ref{e25})-(\ref{e28}) describe the nonlinear dynamics of the large-scale fields in electroconductive medium with temperature inhomogeneity. Connection between the components of a large-scale vortex and magnetic field is carried out by means of the coefficients of the nonlinear hydrodynamic (HD) ${\alpha ^{(1)}}$, ${\alpha ^{(2)}}$ and magnetohydrodynamic (MHD) $\alpha _H^{(1)}$, $\alpha _H^{(2)}$ $\alpha$-effect. Moreover, the coefficients of nonlinear HD and MHD $\alpha$-effect are functions of large-scale velocity fields $\vec W$ and the magnetic field $\vec H$:
	\[{\alpha ^{(1)}} = \frac{{\widetilde {Ra}(1 + P{m^2}\widetilde W_1^2)\left[ {\left( {1 + \Pr )(1 + P{m^{2}}\widetilde W_{1}^{2}\;} \right) + QH_1^2(Pr - Pm)} \right]}}{{2\left[ {{{(1 - Pm\widetilde W_1^2 + QH_1^2)}^2} + \widetilde W_1^2{{(1 + Pm)}^2}} \right]}} \times \]
	\[ \times \left[ {\left( {{{\left( {1 - Pm\widetilde W_1^2 + QH_1^2} \right)}^2} + \widetilde W_1^2{{\left( {1 + Pm} \right)}^2}} \right)\left( {1 + P{r^2}\widetilde W_1^2} \right) + } \right.\]
	\[ + 2Ra\left( {\left( {1 - Pr\widetilde W_1^2} \right)\left( {1 + P{m^2}\widetilde W_1^2} \right) + QH_1^2\left( {1 + Pm\widetilde W_1^2} \right)} \right) + \]
\begin{equation}\label{e29}
  {\left. { + R{a^2}(1 + P{m^2}\widetilde W_1^2)} \right]^{ - 1}},
\end{equation}
	\[{\alpha ^{(2)}} = \frac{{\widetilde {Ra}(1 + P{m^2}\widetilde W_2^2)\left[ {\left( {1 + \Pr {)}(1 + P{m^{2}}\widetilde W_{2}^{2}\;} \right) + QH_2^2(Pr - Pm)} \right]}}{{2\left[ {{{(1 - Pm\widetilde W_2^2 + QH_2^2)}^2} + \widetilde W_2^2{{(1 + Pm)}^2}} \right]}} \times \]
	\[ \times \left[ {\left( {{{\left( {1 - Pm\widetilde W_2^2 + QH_2^2} \right)}^2} + \widetilde W_2^2{{\left( {1 + Pm} \right)}^2}} \right)\left( {1 + P{r^2}\widetilde W_2^2} \right) + } \right.\]
	\[ + 2Ra\left( {\left( {1 - Pr\widetilde W_2^2} \right)\left( {1 + P{m^2}\widetilde W_2^2} \right) + QH_2^2\left( {1 + Pm\widetilde W_2^2} \right)} \right) + \]
\begin{equation}\label{e30}
  {\left. { + R{a^2}(1 + P{m^2}\widetilde W_2^2)} \right]^{ - 1}},
\end{equation}
	\[\alpha _H^{(1)} = \frac{{Pm}}{{({{\left( {1 - Pm\widetilde W_1^2 + QH_1^2} \right)}^2} + \widetilde W_1^2{{\left( {1 + Pm} \right)}^2})}}\left\{ {1 - \frac{}{}} \right.\]
	\[ - Ra\left[ {\left( {1 - Pr\widetilde W_1^2} \right) + \frac{{QH_1^2(1 + PrPm\widetilde W_1^2)}}{{(1 + P{m^2}\widetilde W_1^2)}} + Ra} \right] \times \]
	\[ \times \left[ {\left( {{{\left( {1 - Pm\widetilde W_1^2 + QH_1^2} \right)}^2} + \widetilde W_1^2{{\left( {1 + Pm} \right)}^2}} \right)\frac{{\left( {1 + P{r^2}\widetilde W_1^2} \right)}}{{\left( {1 + P{m^2}\widetilde W_1^2} \right)}} + } \right.\]
\begin{equation}\label{e31}
  \left. {{{\left. { + 2Ra\left[ {\left( {1 - Pr\widetilde W_1^2} \right) + \frac{{QH_1^2\left( {1 + PrPm\widetilde W_1^2} \right)}}{{\left( {1 + P{m^2}\widetilde W_1^2} \right)}}} \right] + R{a^2}} \right]}^{ - 1}}} \right\},
\end{equation}
	\[\alpha _H^{(2)} = \frac{{Pm}}{{({{\left( {1 - Pm\widetilde W_2^2 + QH_2^2} \right)}^2} + \widetilde W_2^2{{\left( {1 + Pm} \right)}^2})}} \times \]
	\[ \times \left\{ {1 - Ra\left[ {\left( {1 - Pr\widetilde W_2^2} \right) + \frac{{QH_2^2(1 + PrPm\widetilde W_2^2)}}{{(1 + P{m^2}\tilde W_2^2)}} + Ra} \right] \times } \right.\]
	\[ \times \left[ {\left( {{{\left( {1 - Pm\widetilde W_2^2 + QH_2^2} \right)}^2} + \widetilde W_2^2{{\left( {1 + Pm} \right)}^2}} \right)\frac{{\left( {1 + P{r^2}\widetilde W_2^2} \right)}}{{\left( {1 + P{m^2}\widetilde W_2^2} \right)}} + } \right.\]
\begin{equation}\label{e32}
  \left. {{{\left. { + 2Ra\left[ {\left( {1 - Pr\widetilde W_2^2} \right) + \frac{{QH_2^2\left( {1 + PrPm\widetilde W_2^2} \right)}}{{\left( {1 + P{m^2}\widetilde W_2^2} \right)}}} \right] + R{a^2}} \right]}^{ - 1}}} \right\}
\end{equation}
\begin{figure}
\centering
   \includegraphics[width=6 cm]{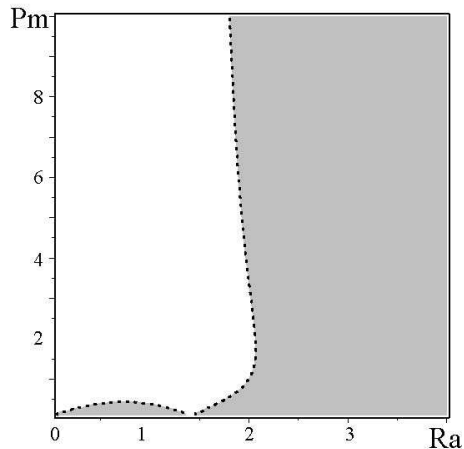}\\
  \caption{ On the the parameter plane $(Ra,Pm)$  the gray area shows the $\gamma  > 1$ and the white shows the region $\gamma  < 1$.}\label{f1}
\end{figure}
Here we use the notation ${\widetilde W_1} = 1 - {W_1}$, ${\widetilde W_2} = 1 - {W_2}$ in order to shorten the recording of formulas (\ref{e29})-(\ref{e32}). Let us note, that HD nonlinear $\alpha$-effect described by the system of equations (\ref{e25})-(\ref{e26}), is possible in presence of  temperature stratification, i.e., Rayleigh number of $Ra \ne 0$ and the external helical forces ${\vec f_0} \ne 0$. On the contrary, a MHD nonlinear $\alpha$-effect occurs,  when there is no heating of $Ra = 0$. In the non electroconductive medium with $\sigma = 0$, the equation (\ref{e25})-(\ref{e26}) coincide with the results of \cite{s18}, \cite{s19}. Like in \cite{s19}, we consider first of all  the stability of the small field disturbances (linear theory), and then examine the possibility of the existence of stationary structures.

\section{Large-scale instability}
\label{fo}

Let us consider the initial stage of development of the perturbations $({W_1},{W_2})$ and $({H_1},{H_2})$. Then, for small values of ${W_1},{W_2}$ and ${H_1},{H_2}$ the equations (\ref{e25})-(\ref{e28}) are linearized and can be reduced to the following system of linear equations:
\begin{equation}\label{e33}
  {\partial _T}{W_1} - \nabla _Z^2{W_1} + \alpha {\nabla _Z}{W_2} = 0
\end{equation}
\begin{equation}\label{e34}
  {\partial _T}{W_2} - \nabla _Z^2{W_2} - \alpha {\nabla _Z}{W_1} = 0
\end{equation}
\begin{equation}\label{e35}
  {\partial _T}{H_1} - \nabla _Z^2{H_1} - {\alpha _H}{\nabla _Z}{H_2} = 0
\end{equation}
\begin{equation}\label{e36}
  {\partial _T}{H_2} - \nabla _Z^2{H_2} + {\alpha _H}{\nabla _Z}{H_1} = 0
\end{equation}
where
\begin{equation}\label{e37}
  \alpha  =  - \frac{{Ra(4 - 2Ra)}}{{{{(4 + R{a^2})}^2}}}\quad \text{for numbers}\quad Pr = 1
\end{equation}
\begin{equation}\label{e38}
  {\alpha _H} = \frac{{2Pm}}{{\left( {1 + P{m^2}} \right)(4 + R{a^2})}}
\end{equation}
From the equations (\ref{e33})-(\ref{e36}) we can see that under small perturbations of fields the self-consistent system of equations (\ref{e25})-(\ref{e28}) splis into two pairs of equations for large-scale field  $\vec W$ and magnetic field $\vec H$ respectively. The first pair of equations (\ref{e33})-(\ref{e34}) are similar to the hydrodynamic equations for $\alpha$-effect \cite{s10}, \cite{s12}, wich generate large-scale vortex structures. The second pair of equations (35)-(36) describes a well-known theory of the dynamo \cite{s26} - \cite{s28}, \cite{s29}, the $\alpha$-effect amplifies  the large-scale magnetic field by the small-scale helical turbulence. In the linear theory considered here, the generation coefficients $\alpha$ and ${\alpha _H}$ do not depend on fields amplitudes but depend only on the characteristics of the medium. For the study of large-scale instability, described by the equations (\ref{e33})-(\ref{e36}), we choose perturbations of velocity $({W_1}, {W_2})$ and magnetic induction $({H_1}, {H_2})$ in the form of flat circular polarized waves:
\begin{equation}\label{e39}
  {W_1} = {A_W}\exp \left( {\Gamma t} \right)\;sinKZ,\;{W_2} = {B_W}\exp \left( {\Gamma t} \right)\;\cos KZ
\end{equation}
\begin{equation}\label{e40}
  {H_1} = {A_H}\exp \left( {\Gamma t} \right)\;sinKZ,\;{H_2} = {B_H}\exp \left( {\Gamma t} \right)\;\cos KZ
\end{equation}
\begin{figure}
  \centering
  \includegraphics[height=5 cm]{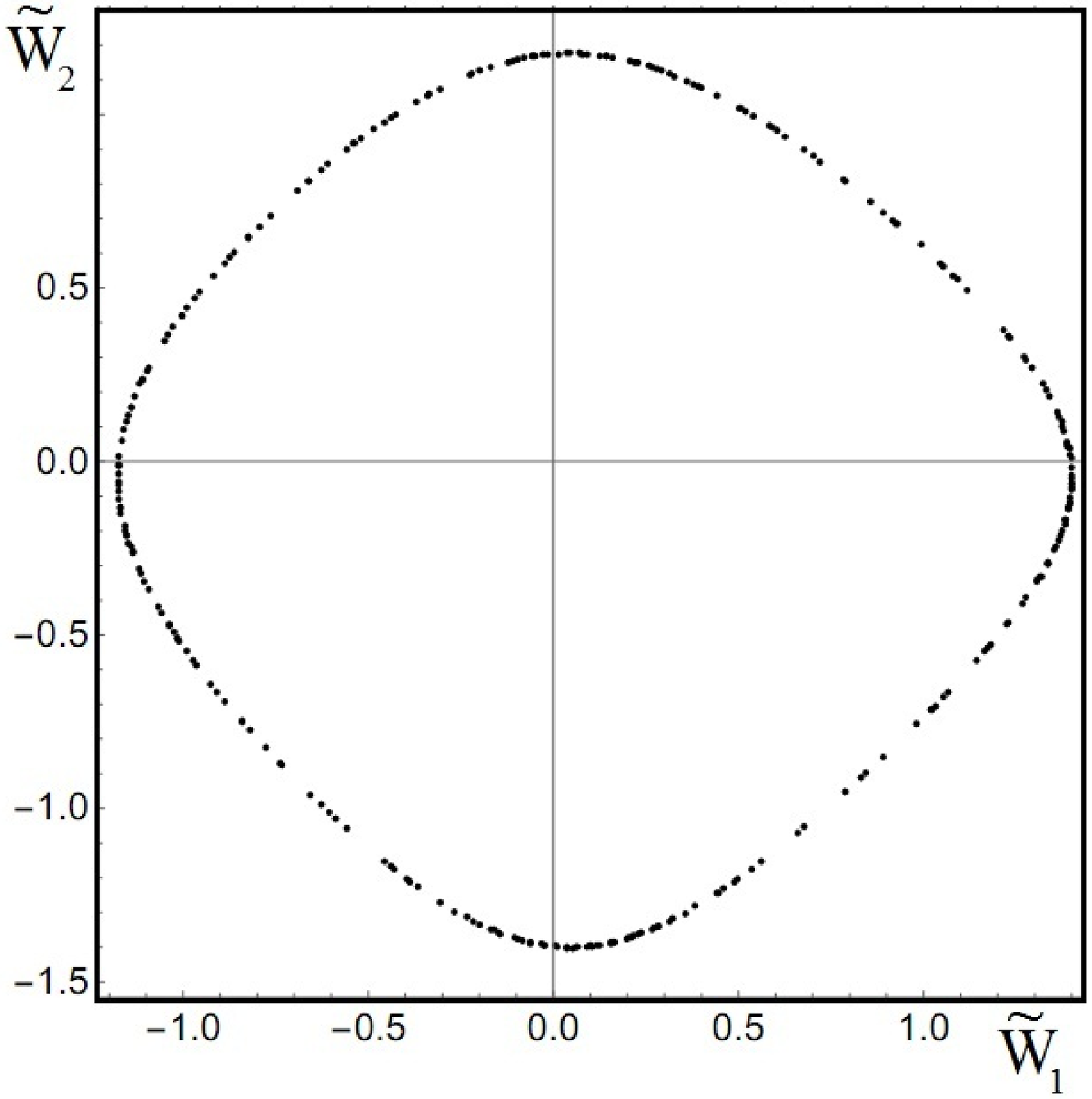}
  \includegraphics[height=5 cm]{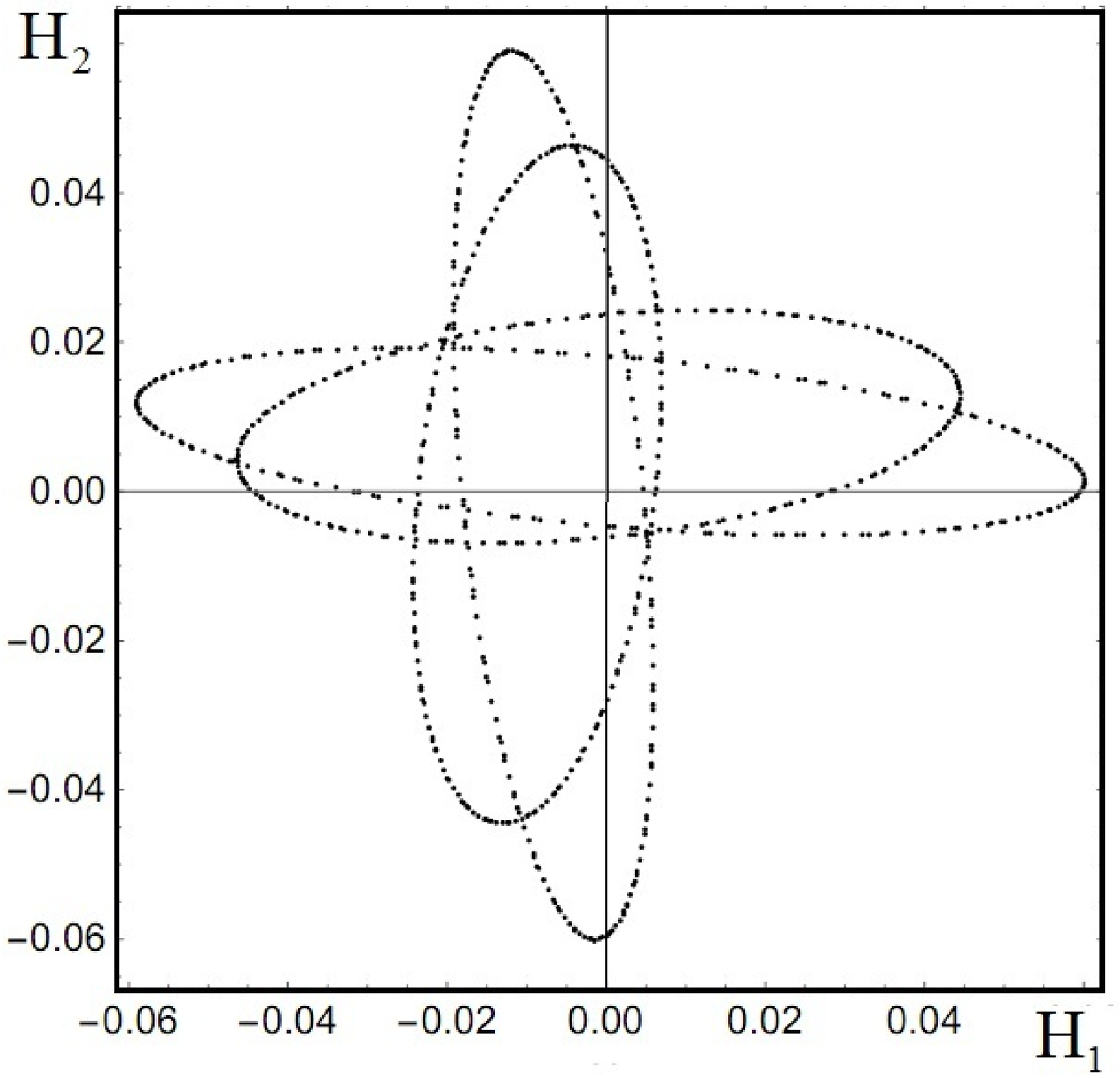}\\
  \includegraphics[height=5 cm]{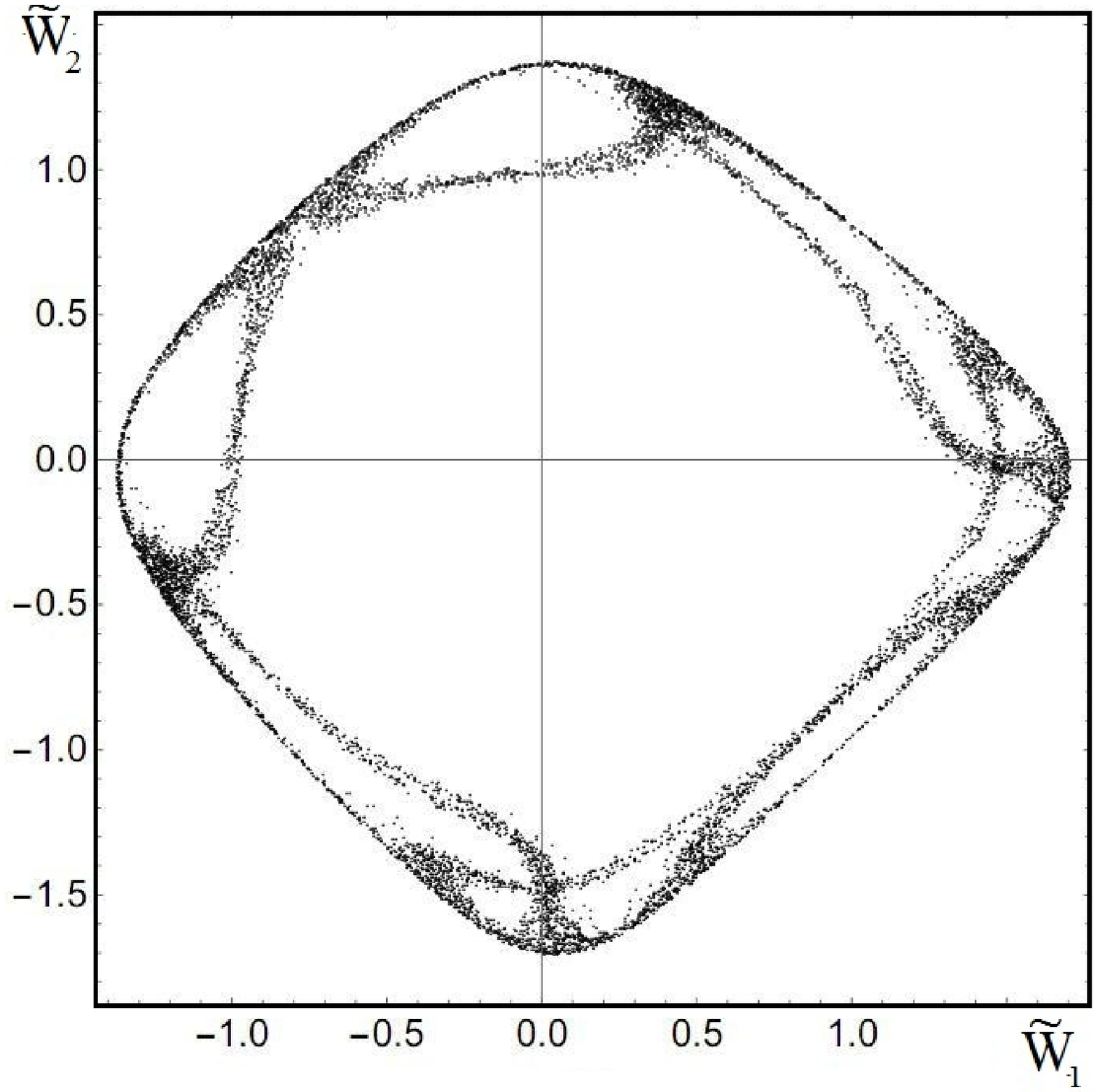}
  \includegraphics[height=5 cm]{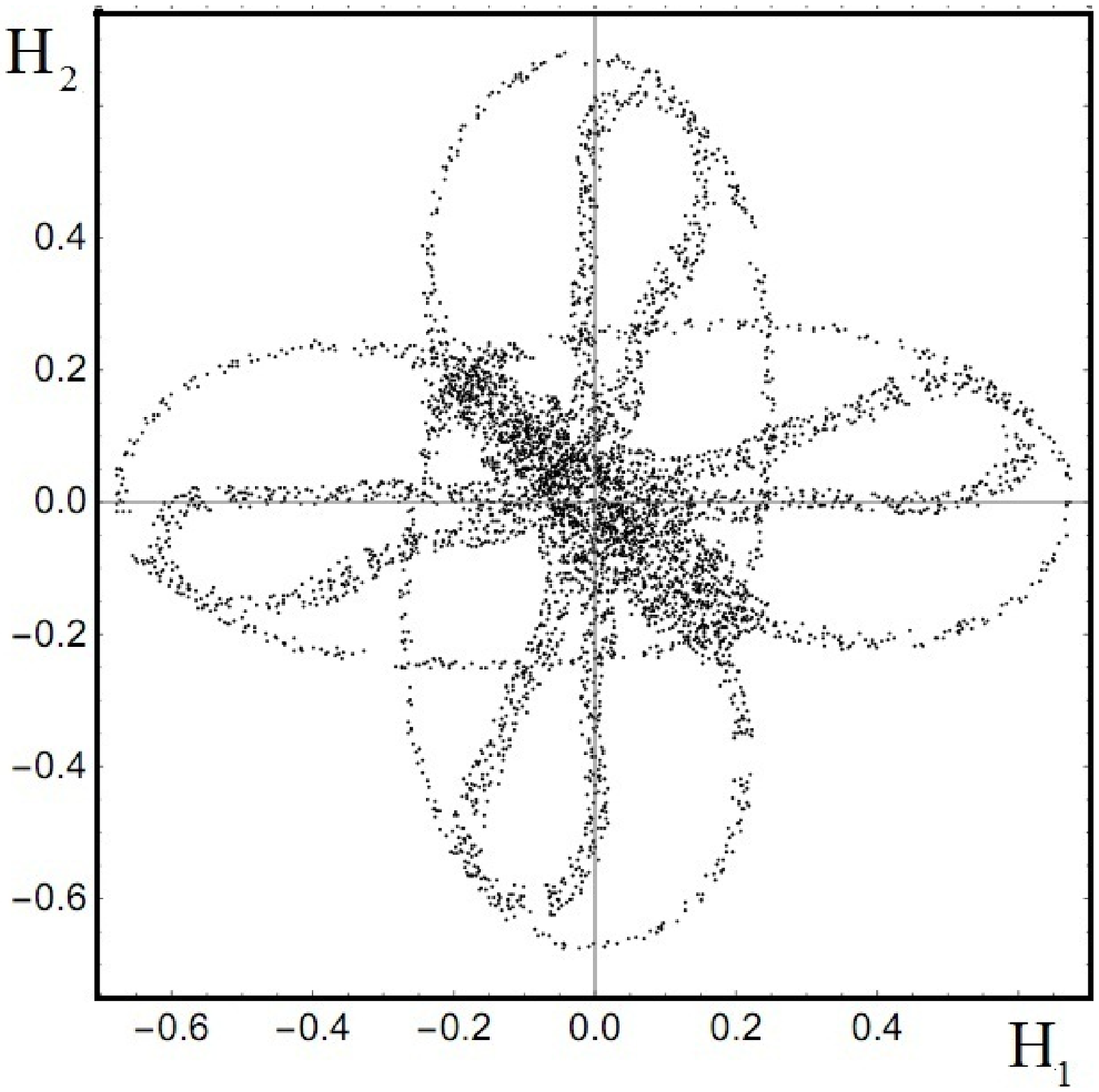}\\
  \caption{Poincare section for two trajectories. The upper, for the trajectory with the initial conditions ${\widetilde W_1}(0) = 0.8$, ${\widetilde W_2}(0) = 0.8$, ${H_1}(0) = 0.01$, ${H_2}(0) = 0.01$ and the lower with the initial conditions ${\widetilde W_1}(0) = 0.9$, ${\widetilde W_2}(0) = 0.9$, ${H_1}(0) = 0.01$, ${H_2}(0) = 0.01$. For the left figures, the cutting plane is stretched to single coordinate vectors of the velocity, and for the right ones to coordinate vectors of the magnetic field. One can see that the trajectory corresponding to the upper figures is wounded on tori. These trajectories are regular. The lower figures show stochastic layers. The corresponding chaotic trajectories belong to them.}\label{f2}
\end{figure}

Substituting (\ref{e39}) in the system of equations (\ref{e33})-(\ref{e34}), and (\ref{e40}) in (\ref{e35})-(\ref{e36}), we obtain two instability increments:
\begin{equation}\label{e41}
  {\Gamma _1} =  \pm \alpha K - {K^2}
\end{equation}
\begin{equation}\label{e42}
  {\Gamma _2} =  \pm {\alpha _H}K - {K^2}P{m^{ - 1}}
\end{equation}
The growing solution with the first increment, describes the generation of helical vortex structures of Beltrami type. The maximum of increment ${\Gamma _{1max}} = \frac{{{\alpha ^2}}}{4}$ is reached at ${K_{max}} = \frac{\alpha }{2}$. In the same way, we find from formula (\ref{e42}) the maximum increment of large-scale magnetic field generation ${\Gamma _{2max}} = \frac{{\alpha _H^2}}{4}Pm$ at ${K_{max}} = \frac{{{\alpha _H}}}{2}Pm$.

If the external force has zero helicity ${\vec F_0}rot{\vec F_0} = 0$, the two $\alpha$-effects disappear: $\alpha = 0$, ${\alpha _H} = 0$. In addition, when the temperature gradient is vanishing, the hydrodynamic $\alpha$-effect also disappears, but MHD $\alpha$-effect remains. So in these conditions, the magnetic field continues to grow. In order to understand for which Rayleigh number $Ra$ are generated large-scale vortices or magnetic disturbances it is convenient to introduce the coefficient of relative growth rate of perturbations $\gamma  = {\Gamma _{1\max }}/{\Gamma _{2\max }}$:
\begin{equation}\label{e43}
  \gamma  = \frac{{{\Gamma _{1max}}}}{{{\Gamma _{2max}}}} = \frac{{R{a^2}{{(4 - 2Ra)}^2}{{(1 + P{m^2})}^2}}}{{4P{m^3}{{(4 + R{a^2})}^2}}}
\end{equation}
From this we can see that at small magnetic Prandtl $Pm \ll 1$ and Rayleigh numbers of $Ra > 2$ there is the most effective generation of large-scale vortex disturbances. The small magnetic Prandtl numbers of $Pm$ can be due to the low electrical conductivity $(\sigma  \to 0)$ medium and very low kinematic viscosity $(\nu  \to 0)$ of medium. Now consider the case of low electroconductivity medium $(\sigma  \to 0)$ with $\nu  \ne 0$, then the magnetic number of Prandtl and Chandrasekhar number are small : $Pm \to 0$, $Q \to 0$. In this medium, the generation of large-scale magnetic field is not effective ${H_{1,2}} \ll 1$ because the coefficient $\gamma  \gg  1$, and under the influence of  external helical small-scale force  the generation of  large-scale vortex structures is possible in the convective medium which are described in detail in works \cite{s18}, \cite{s19}. We turn now to the case of an electrically conductive medium, i.e., when the magnetic number of Prandtl is non zero $Pm \ne 0$. From the numerical analysis of the formula (\ref{e43}) (see fig.\ref{f1}) it follows  that the generation of vortex disturbances is most effecient when Rayleigh numbers are $Ra > 2$. White area in the figure\ref{f1} shows the area of the preferential generation of magnetic disturbances. We can state that the magnetic field generation is most effective for Rayleigh numbers in the interval $Ra \in [0,2]$. The linear theory is not correct when  the large-scale perturbations are strong enough. Therefore it is necessary to take into account the nonlinear  effects.

\section{Stationary nonlinear magnetic structures}
\label{fi}

We turn now to a discussion of the nonlinear stage. Taking into account the dependence of the right sides of the system of nonlinear equations of $\vec W$, $\vec H$, one would expect that with the growth of perturbations the nonlinear coefficients of ${\alpha ^{(1)}}$, ${\alpha ^{(2)}}$, $\alpha _H^{(1)}$, $\alpha _H^{(2)}$ decrease and the instability is stabilized. As a result the nonlinear stationary structures are formed. To describe these structures let us examine the nonlinear system of equations (25) - (28) in the stationary case, taking ${\partial _T}{W_1} = {\partial _T}{W_2} = {\partial _T}{H_1} = {\partial _T}{H_2} = 0$.  Integrating these equations on $Z$ we obtain:
\begin{equation}\label{e44}
  \frac{{d{{\widetilde W}_1}}}{{dZ}} = {\alpha ^{(2)}}{\widetilde W_2}\left( {1 - \frac{{QH_2^2Pm}}{{1 + P{m^2}\widetilde W_2^2}}} \right) + {C_1}
\end{equation}
\begin{equation}\label{e45}
  \frac{{d{{\widetilde W}_2}}}{{dZ}} =  - {\alpha ^{(1)}}{\widetilde W_1}\left( {1 - \frac{{QH_1^2Pm}}{{1 + P{m^2}\widetilde W_1^2}}} \right) + {C_2}
\end{equation}
\begin{equation}\label{e46}
  \frac{1}{{Pm}}\frac{{d{H_1}}}{{dZ}} =  - \alpha _H^{(2)}{H_2} + {C_3}
\end{equation}
\begin{equation}\label{e47}
  \frac{1}{{Pm}}\frac{{d{H_2}}}{{dZ}} = \alpha _H^{(1)}{H_1} + {C_4}
\end{equation}
 Here ${C_1}$, ${C_2}$, ${C_3}$, ${C_4}$ are arbitrary constants of integration. Equations (\ref{e44}) - (\ref{e47}) present the nonlinear dynamical system in four-dimensional phase space. It can be proved that this system of equations is conservative. However to find the Hamiltonian of this nonlinear system is technically  bulky task. Even if it exists, it can only be obtained in quadratures and the execution of integration removes it beyond the class of elementary functions.
\begin{figure}
  \centering
  \includegraphics[height=5 cm]{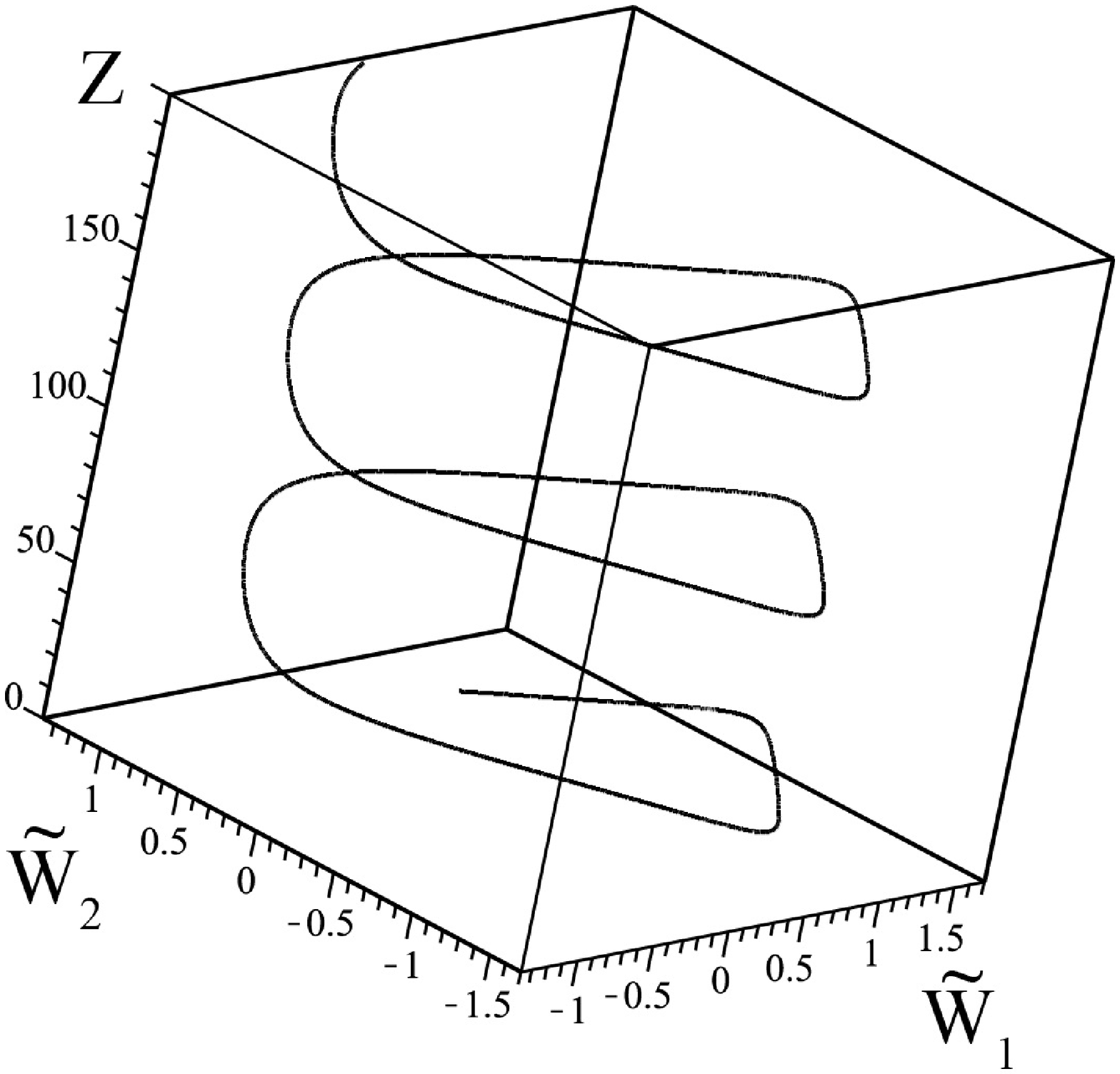}
  \includegraphics[height=5 cm]{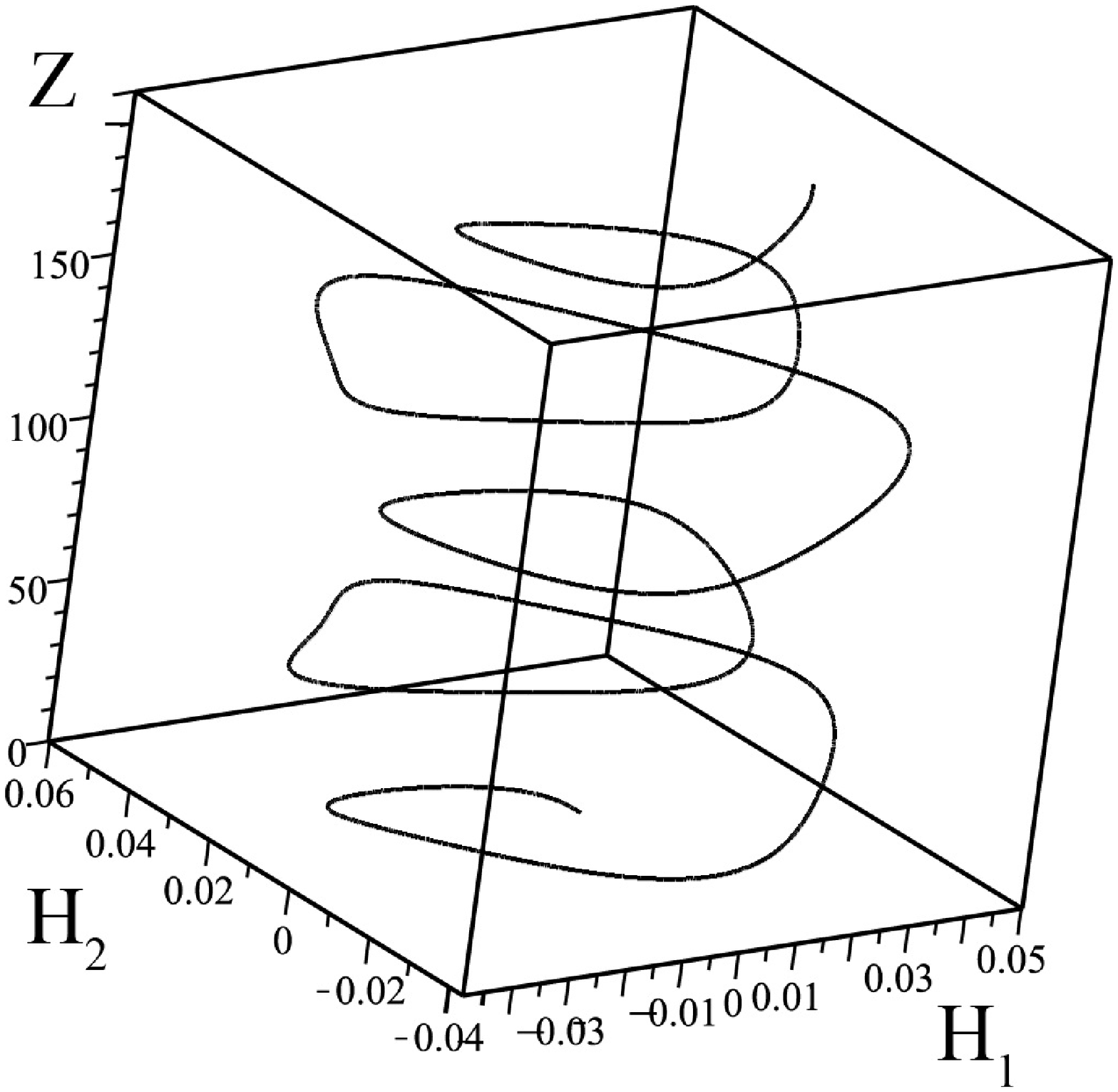}\\
  \includegraphics[height=5 cm]{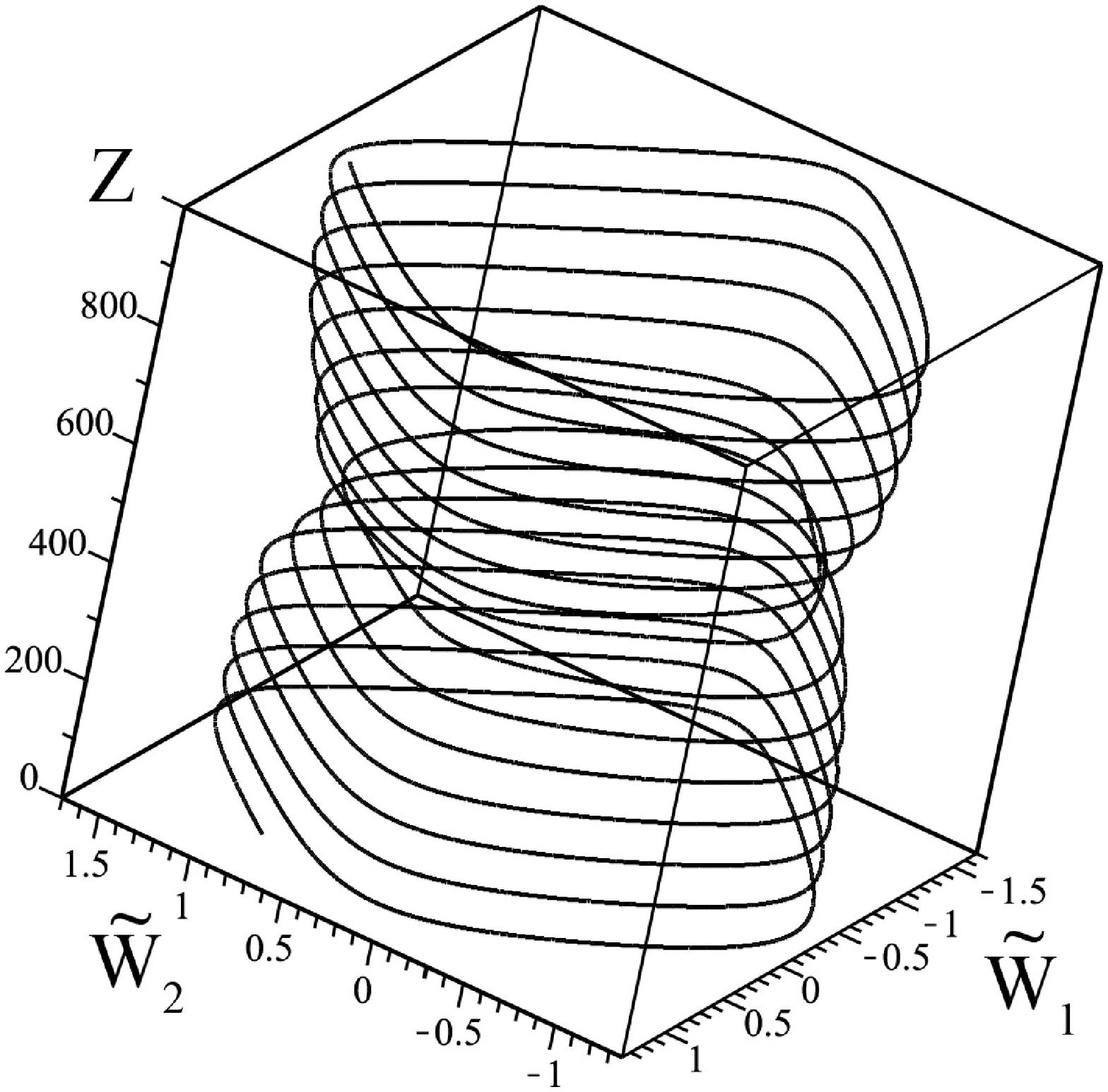}
  \includegraphics[height=5 cm]{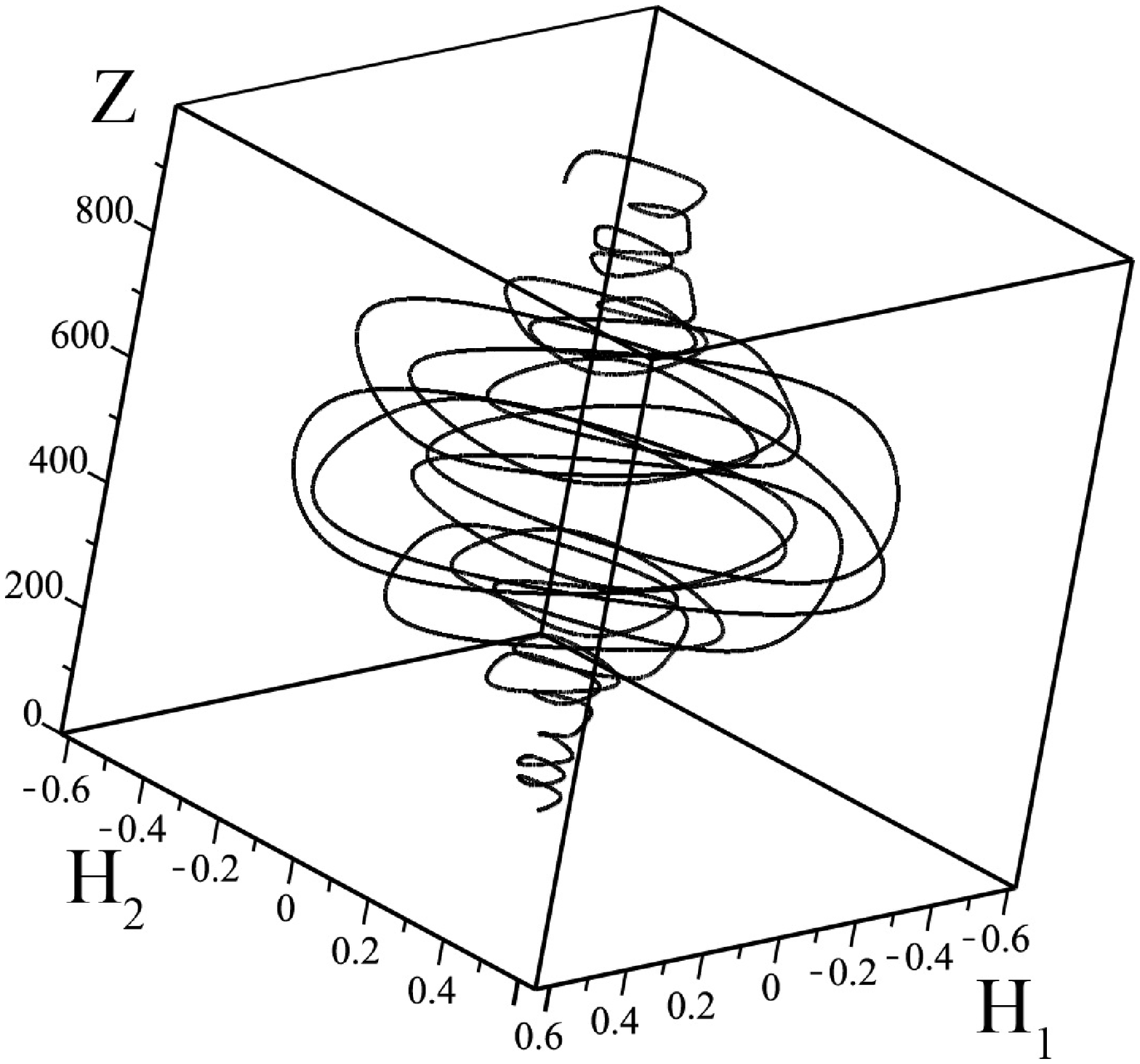}\\
  \caption{The upper part shows the velocity and magnetic field dependence on the height for the numerical solution of equations (\ref{e44}) - (\ref{e47}) with the initial conditions: ${\widetilde W_1}(0) = 0.8$, ${\widetilde W_2}(0) = 0.8$, ${H_1}(0) = 0.01$, ${H_2}(0) = 0.01$. This dependence corresponds to regular motion of the Poincare section shown on top of Fig. \ref{f2}. Below a similar dependence for the numerical solution of equations ((44)) - ((47)) with the initial conditions: ${\widetilde W_1}(0) = 0.9$, ${\widetilde W_2}(0) = 0.9$, ${H_1}(0) = 0.01$, ${H_2}(0) = 0.01$. This chaotic dependence corresponds to Poincare sections in Fig.\ref{f2}  shown at the bottom.}\label{f3}
\end{figure}

In general case this conservative nonlinear system of four coupled equations has no attractors in phase space. The high dimensionality of the phase space and a large number of parameters in the system make very diffucult the full qualitative analysis of this system. One of the common features is the invariance of the system under the transformation $({\widetilde W_1},{\widetilde W_2},{H_1},{H_2}) \to ( - {\widetilde W_1}, - {\widetilde W_2}, - {H_1}, - {H_2})$. With zero constant ${C_1} = {C_2} = {C_3} = {C_4} = 0$ in the phase space a fixed point exists at the origin of the phase space.  In the phase space of such a system  the presence of resonant and nonresonant tori can be expected. In its turn this means the existence of chaotic stationary structures of hydrodynamic and magnetic fields. To prove the existence of these stationary trajectories let us consider the Poincare section of the trajectories in the phase space. Fig.\ref{f2} shows examples of the cross-sections obtained numerically for the dimensionless parameters $Q = Pm = Pr = 1$, $Ra = 2$ and the constants ${C_1} = {C_2} = 0.01$, ${C_3} = {C_4} = 0.001$. The upper part of Fig.\ref{f2} demonstrates the Poincare section of regular trajectory for the velocity and magnetic fields. The structure of the chaotic layer, which belongs to the selected path is clearly visible on the lower section. The presence of such chaotic trajectories implies the existence of a stationary random structure of the velocity and magnetic fields. The chaotic change of velocity direction of magnetic fields according to the height $Z$ is typical for these trajectories. Thus this set of equations has  stationary chaotic solutions. Fig.\ref{f3} shows the dependence of stationary large-scale fields on the height of $Z$, which was obtained numerically for the initial conditions, which correspond to the Poincare sections in Fig.\ref{e2}. These figures show also the appearance of stationary random solutions for magnetic and vortex fields. It should be noted that the structure of the magnetic field at the bottom of Fig.\ref{f3} demonstrates with the increasing of height the intermittency structure.  In the numerical solution of equations (\ref{e44})-(\ref{e47}) occur chaotic structures observed with the increase of the initial velocity amplitudes. For small initial velocities and magnetic fields regular trajectories are typical. With increasing of velocity amplitude,  above a certain critical value the chaotic trajectories appear.  With increasing of the initial velocity the part of the space occupied by the chaotic trajectories points grows on the Poincare sections. Under these conditions, typical solutions become chaotic.

Let us now consider in more detail which nonlinear structure may occur in the convective electrically conductive turbulent medium, in some limit cases. Let us suppose that the generation of large-scale vortex disturbances in the convective medium is not yet in the stationary mode, but large-scale perturbations of the magnetic field are already reached their saturation on  stationary level. Then the influence of the small amplitude of large-scale vortex disturbances ${W_{1,2}} \ll 1$ on the evolution of large-scale magnetic fields can be neglected. As a result, from the equations of the nonlinear dynamo (\ref{e25}) - (\ref{e28}), we obtain the equations for the evolution of the large-scale stationary magnetic field:
\begin{equation}\label{e48}
  \frac{{d{H_1}}}{{dZ}} =  - \frac{{{H_2}({Q^2}H_2^4 + RaQH_2^2 + 4)}}{{({Q^2}H_2^4 + 4)({Q^2}H_2^4 + 2RaQH_2^2 + R{a^2} + 4)}} + {C_3}
\end{equation}
\begin{equation}\label{e49}
  \frac{{d{H_2}}}{{dZ}} = \frac{{{H_1}({Q^2}H_1^4 + RaQH_1^2 + 4)}}{{({Q^2}H_1^4 + 4)({Q^2}H_1^4 + 2RaQH_1^2 + R{a^2} + 4)}} + {C_4}
\end{equation}
Here, to simplify the calculation we use the Prandtl numbers: $Pr = Pm = 1$.
\begin{figure}
  \centering
  \includegraphics[height=4 cm]{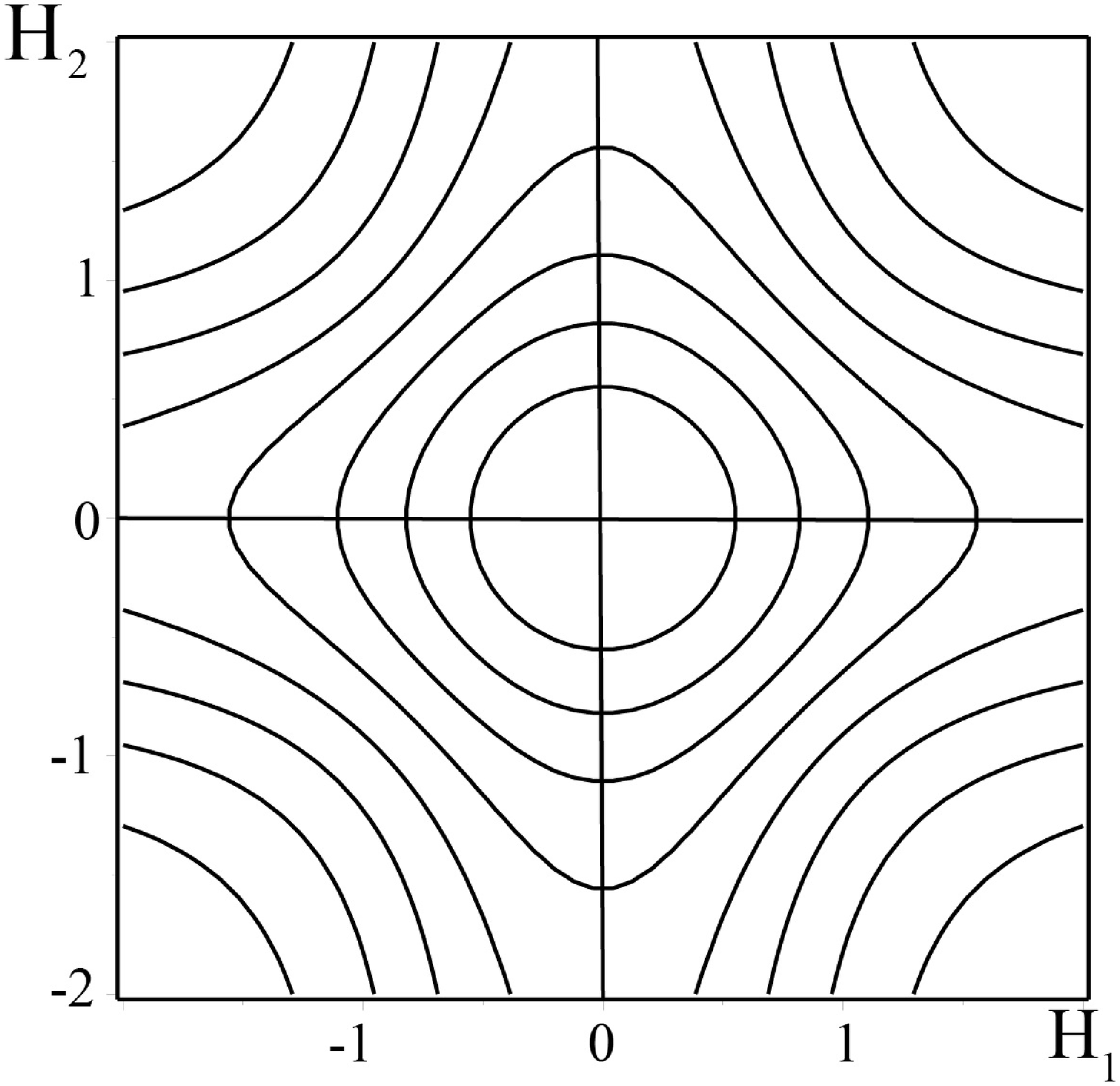}
  \includegraphics[height=4 cm]{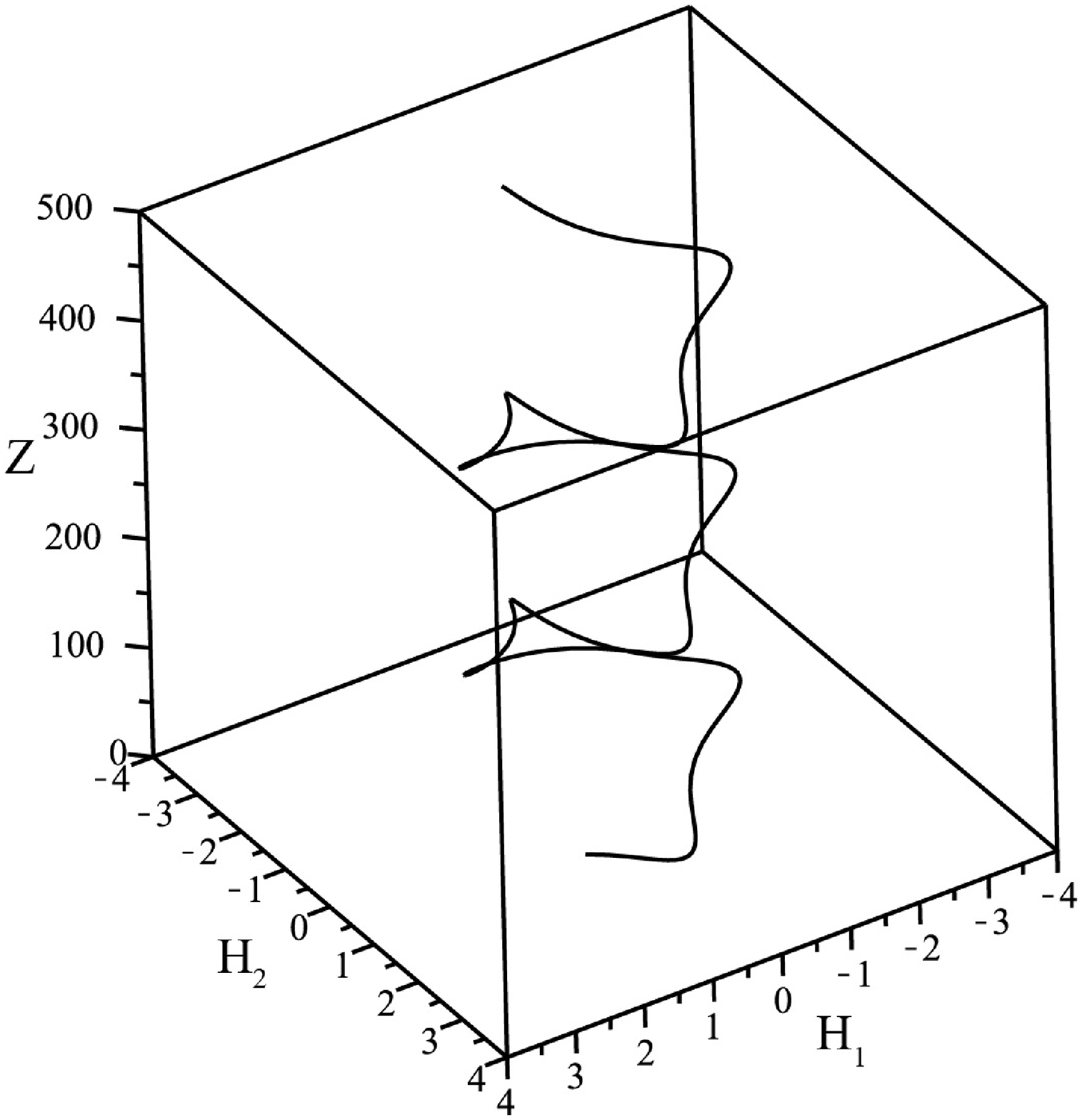}\\
  \caption{ On the left the phase portrait of Hamiltonian equations (\ref{e51}) at ${C_3} = {C_4} = 0$; on the right, the solution corresponding to the  nonlinear wave.}\label{f4}
\end{figure}
Equation (\ref{e48})-(\ref{e49}) can be written in the Hamiltonian form:
\begin{equation}\label{e50}
  \frac{{d\widetilde X}}{{dt}} =  - \frac{{d{\cal H}}}{{d\widetilde P}},\;\;\frac{{d\widetilde P}}{{dt}} = \frac{{d{\cal H}}}{{d\widetilde X}}
\end{equation}
where we have introduced new variables $\widetilde X = \sqrt Q {H_1}$, $\widetilde P = \sqrt Q {H_2}$, $t = Z$. The Hamiltonian of the magnetic field ${\cal H}$ is of the form:
\begin{equation}\label{e51}
  {\cal H} = U(\widetilde P) + U(\widetilde X) - {\widetilde C_3}P + {\widetilde C_4}X + {C_5}
\end{equation}
where the function $U(y)$ is equal to:
\[ U(y) = \frac{{Ra}}{{4(R{a^2} + 16)}}\ln \left( {\frac{{{y^4} + 4}}{{{y^4} + 4 + 2Ra{y^2} + R{a^2}}}} \right)+\]
\begin{equation}\label{e52}
  + \frac{2}{{R{a^2} + 16}}\left( {arctg\left( {\frac{{{y^2}}}{2}} \right) + arctg\left( {\frac{{{y^2} + Ra}}{2}} \right)} \right)
\end{equation}
The constant ${\widetilde C_3}$ and ${\widetilde C_4}$, respectively, are: ${\widetilde C_3} = \sqrt Q {C_3}$, ${\widetilde C_4} = \sqrt Q {C_4}$. We examine now the the types of stationary magnetic structures described by equations (\ref{e50}). First of all we consider in detail the limit case $Ra \to 0$. Then the Hamiltonian (\ref{e51}) takes the following form:
\begin{equation}\label{e53}
  {\cal H} \to {\cal H}' = \frac{1}{{4Q}}arctg\left( {\frac{{QH_1^2}}{2}} \right) + \frac{1}{{4Q}}arctg\left( {\frac{{QH_2^2}}{2}} \right) + {C_4}{H_1} - {C_3}{H_2} + {\widetilde C_5}
\end{equation}

The Hamiltonian equations (\ref{e50}) mean that in the phase space only fixed points of two types can be observed: elliptical and hyperbolic fixed points. At zero values of the constants ${C_3} = {C_4} = 0$ it is possible to construct the phase portrait of Hamiltonian (\ref{e53}) from which it is clear that there is only one elliptic point (see Fig.\ref{f4}). Around the elliptic points  are observed the nonlinear waves only. The numerical stationary solution of equations (\ref{e48})-(\ref{e49}) in the limit of $Ra \to 0$ corresponds to a nonlinear wave of finite amplitude and is shown in Fig.\ref{f4}. Calculating the maximum and minimum of $\left( {\frac{{{H_{1,2}}}}{{{Q^2}H_{1,2}^4 + 4}}} \right)$ we find the area of the parameter $({C_3},{C_4})$ change defined by the inequalities:
	\[ - \chi  < {C_3} < \chi \]

	\[ - \chi  < {C_4} < \chi \]
where $\chi  = \frac{{{3^{3/4}}}}{{16}}\sqrt {\frac{2}{Q}}$ depends on the choice of the parameter $Q$.
\begin{figure}
  \centering
  \includegraphics[width=5 cm]{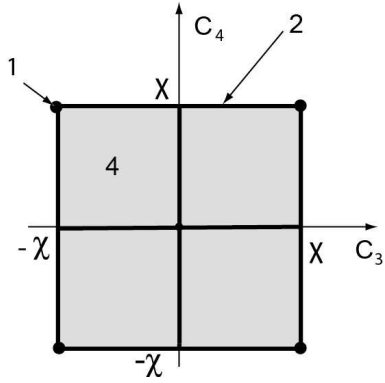}\\
  \caption{On the parameter plane $({C_3},{C_4})$ gray area shows the presence of four fixed points. For constants $({C_3} = 0,{C_4} = 0)$, there is only one fixed point. The bold lines mark the values of the constants which correspond to the two fixed points on phase portrait. The vertices of the square area correspond to the values at which  exist one fixed point. For values outside this area fixed points are absent.}\label{f5}
\end{figure}

\begin{figure}
  \centering
  \includegraphics[height=4 cm]{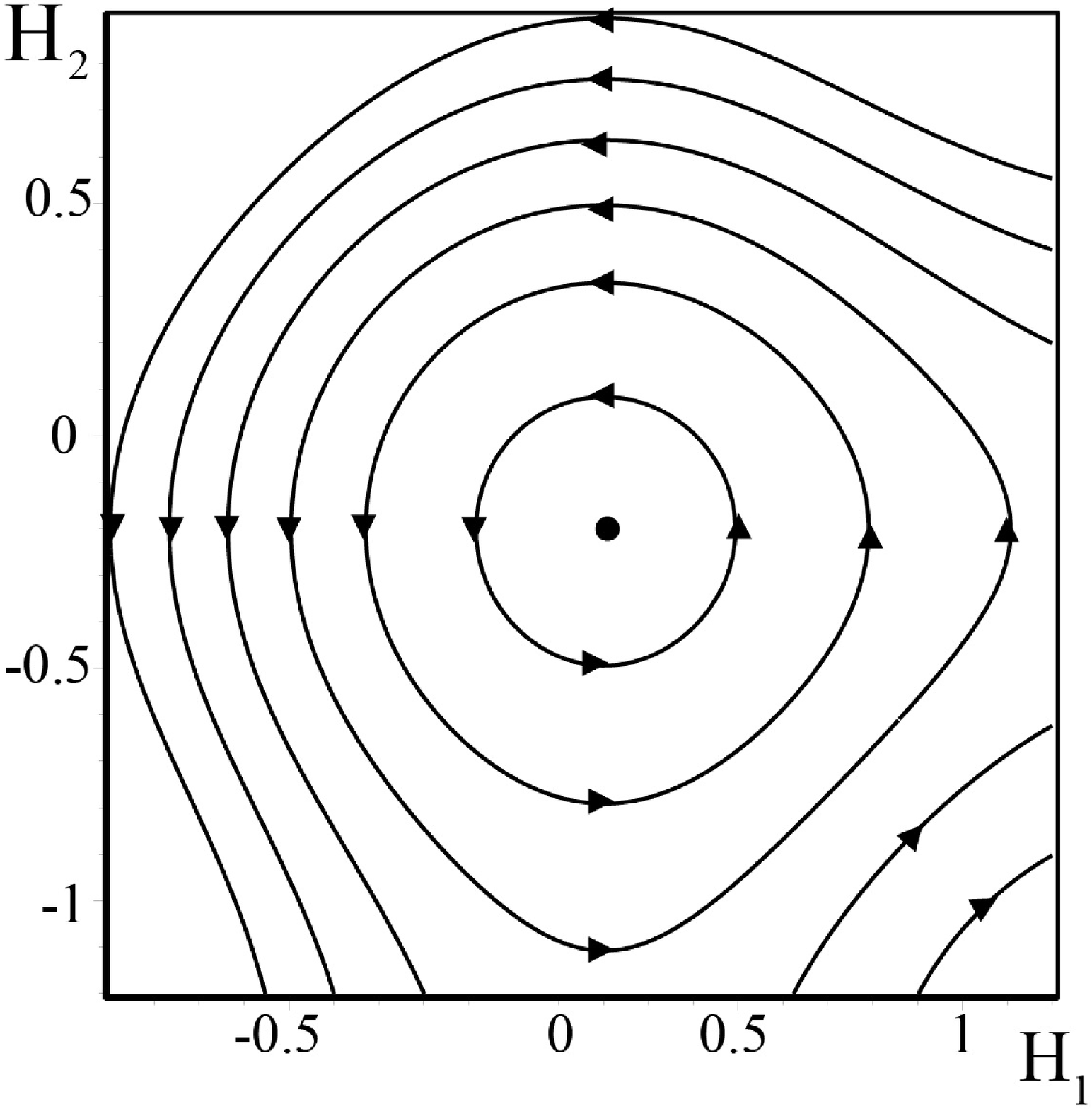}
  \includegraphics[height=4 cm]{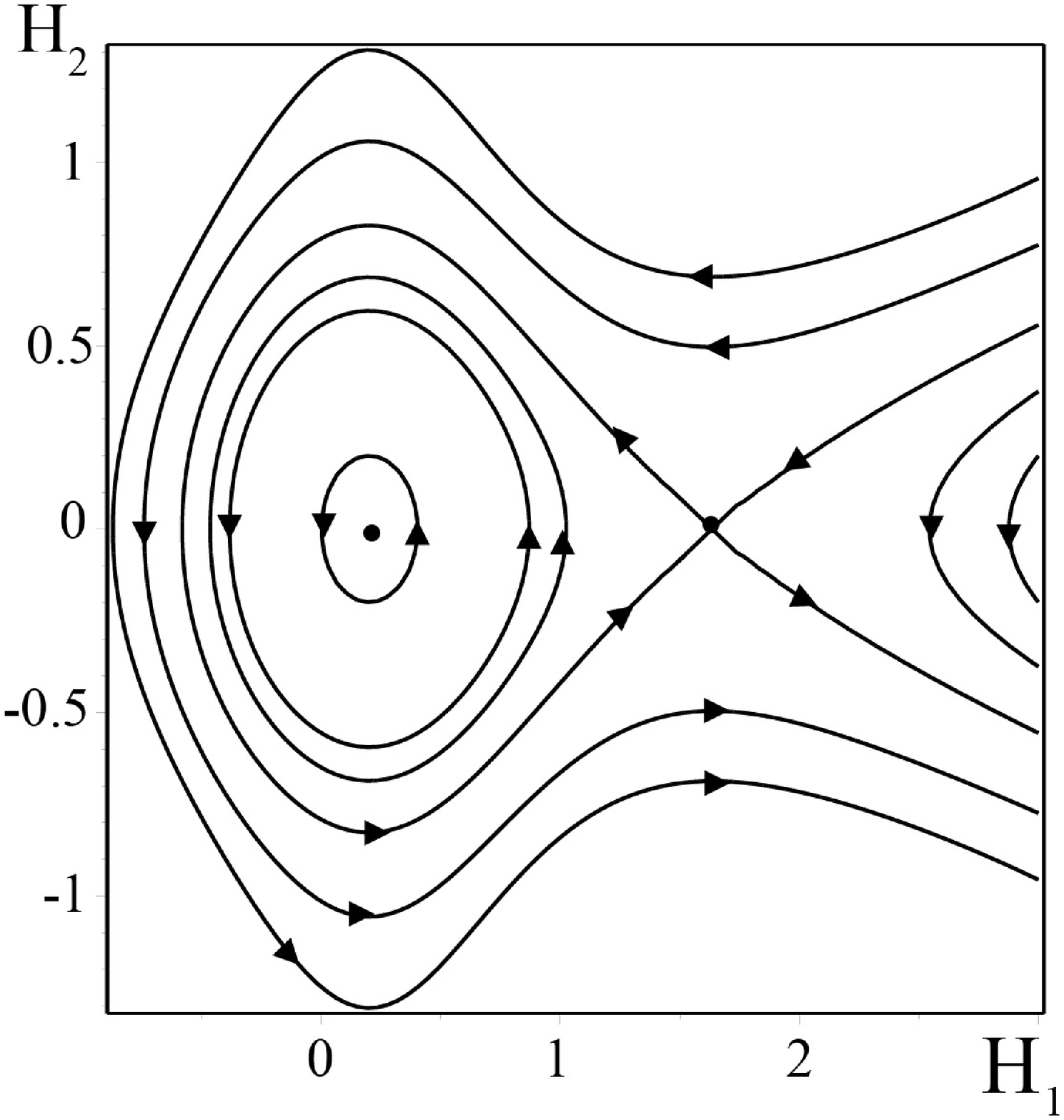}
  \includegraphics[height=4 cm]{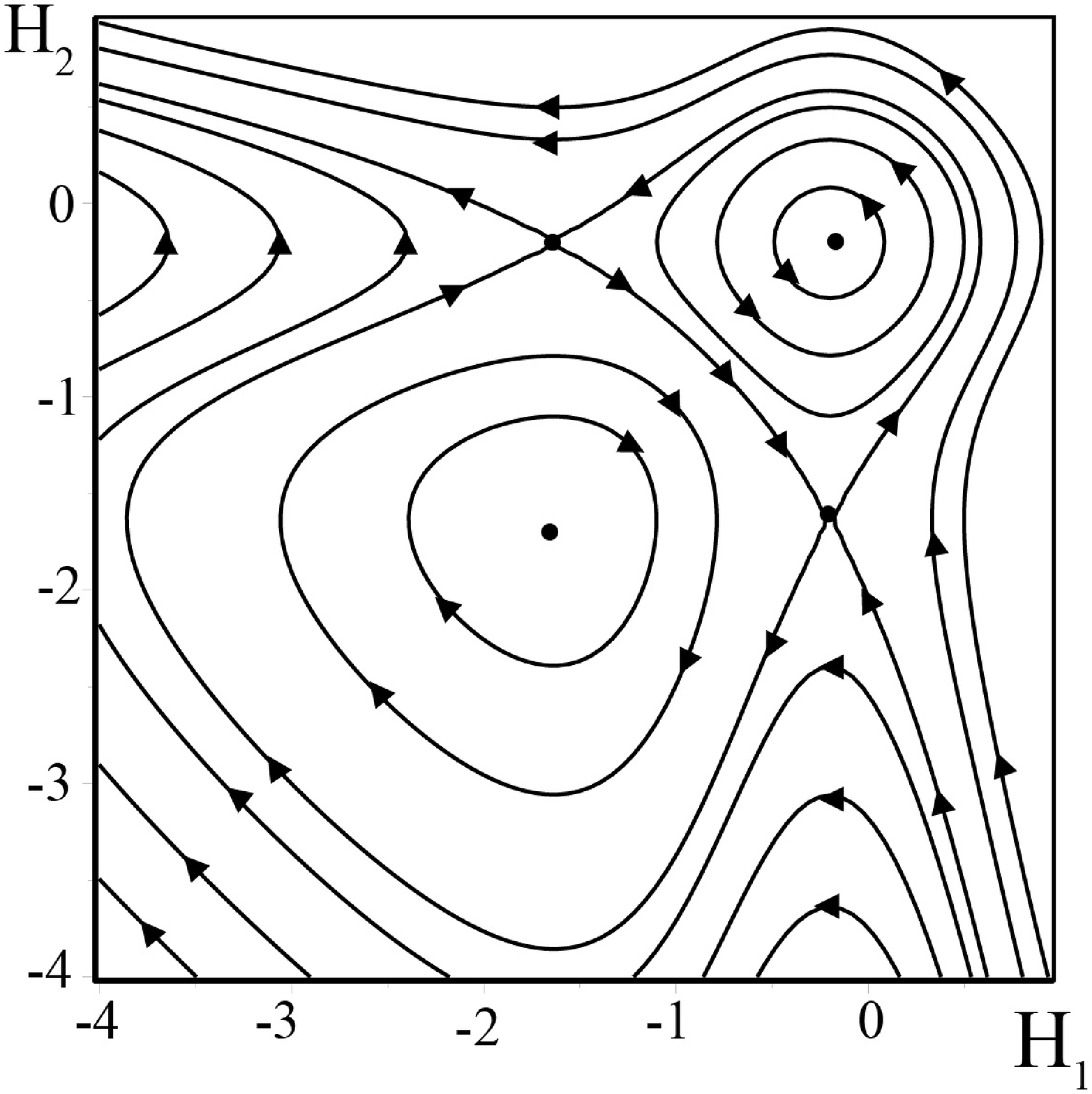}\\
  \caption{On the left the phase portrait 	for the fixed points located at vertices of the square area of parameters $({C_3},{C_4})$; the center-phase portrait for the fixed points on the boundary of the parameters $({C_3},{C_4})$, fixed points when $({C_3} = 0,{C_4} =  \pm \chi )$; $({C_3} =  \pm \chi ,{C_4} = 0)$; on the right the phase portrait for the fixed points inside the square area of parameters $({C_3},{C_4})$.}\label{f6}
\end{figure}

\begin{figure}
  \centering
  \includegraphics[height=4 cm]{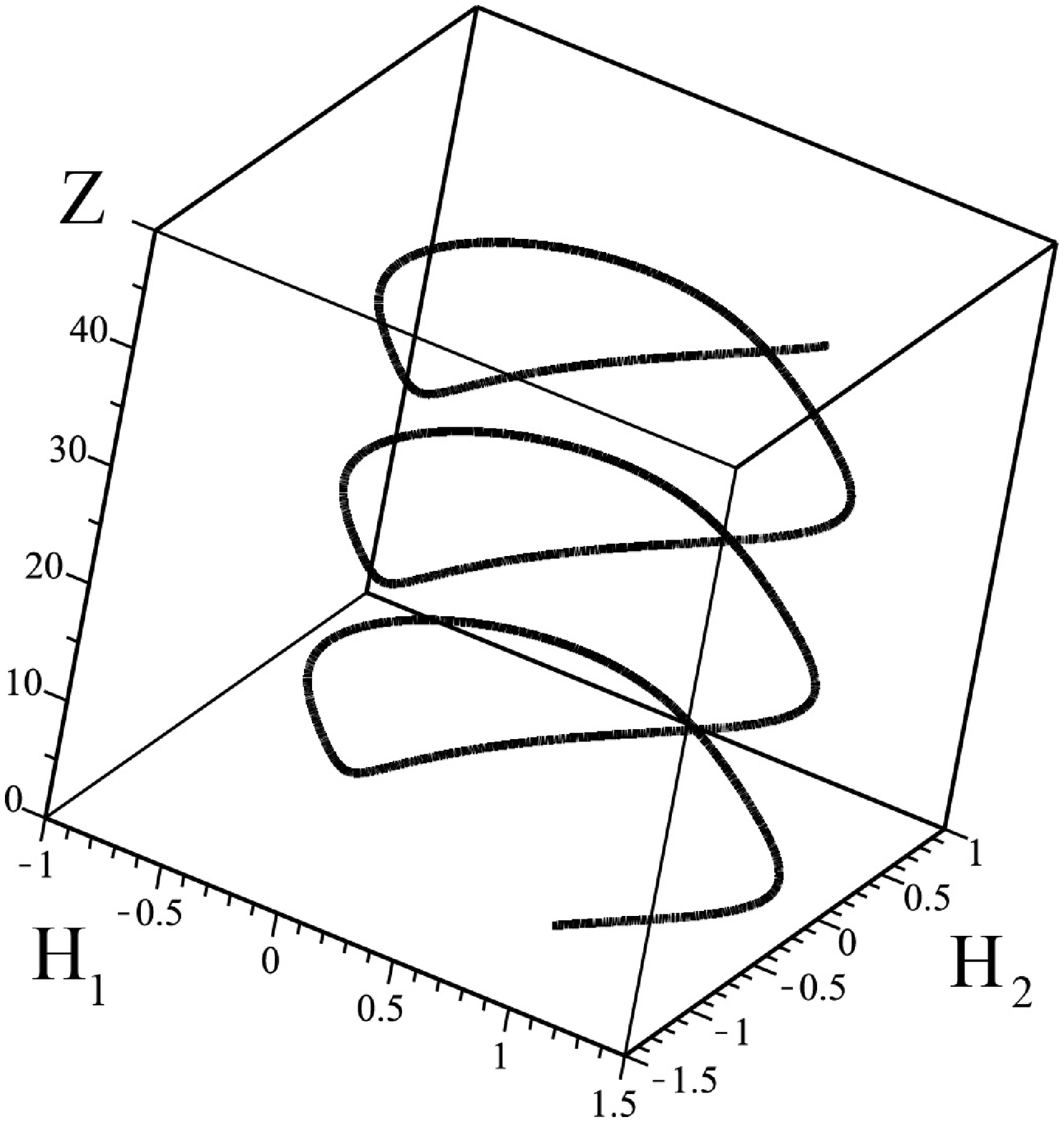}
  \includegraphics[height=4 cm]{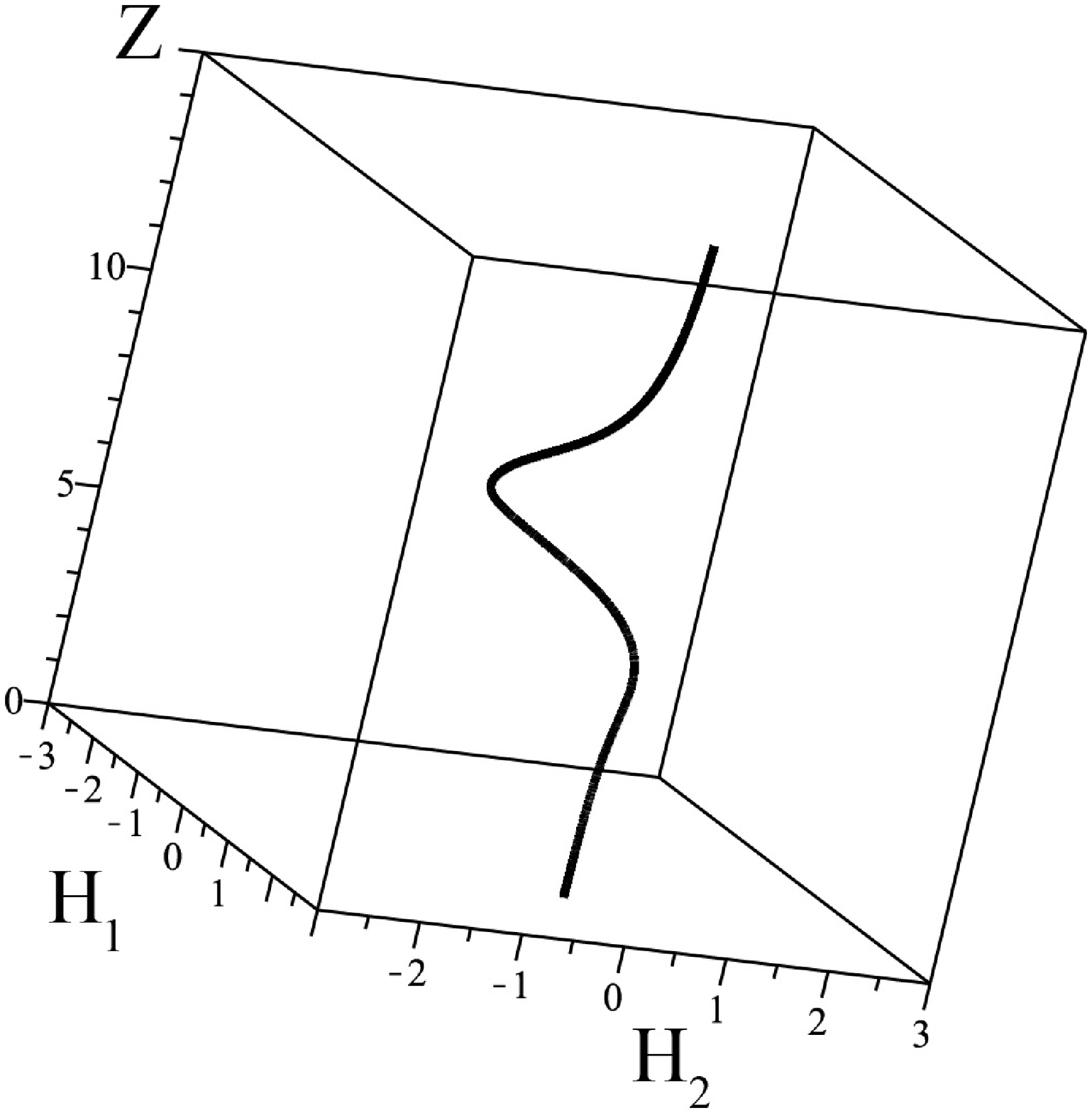}
  \includegraphics[height=4 cm]{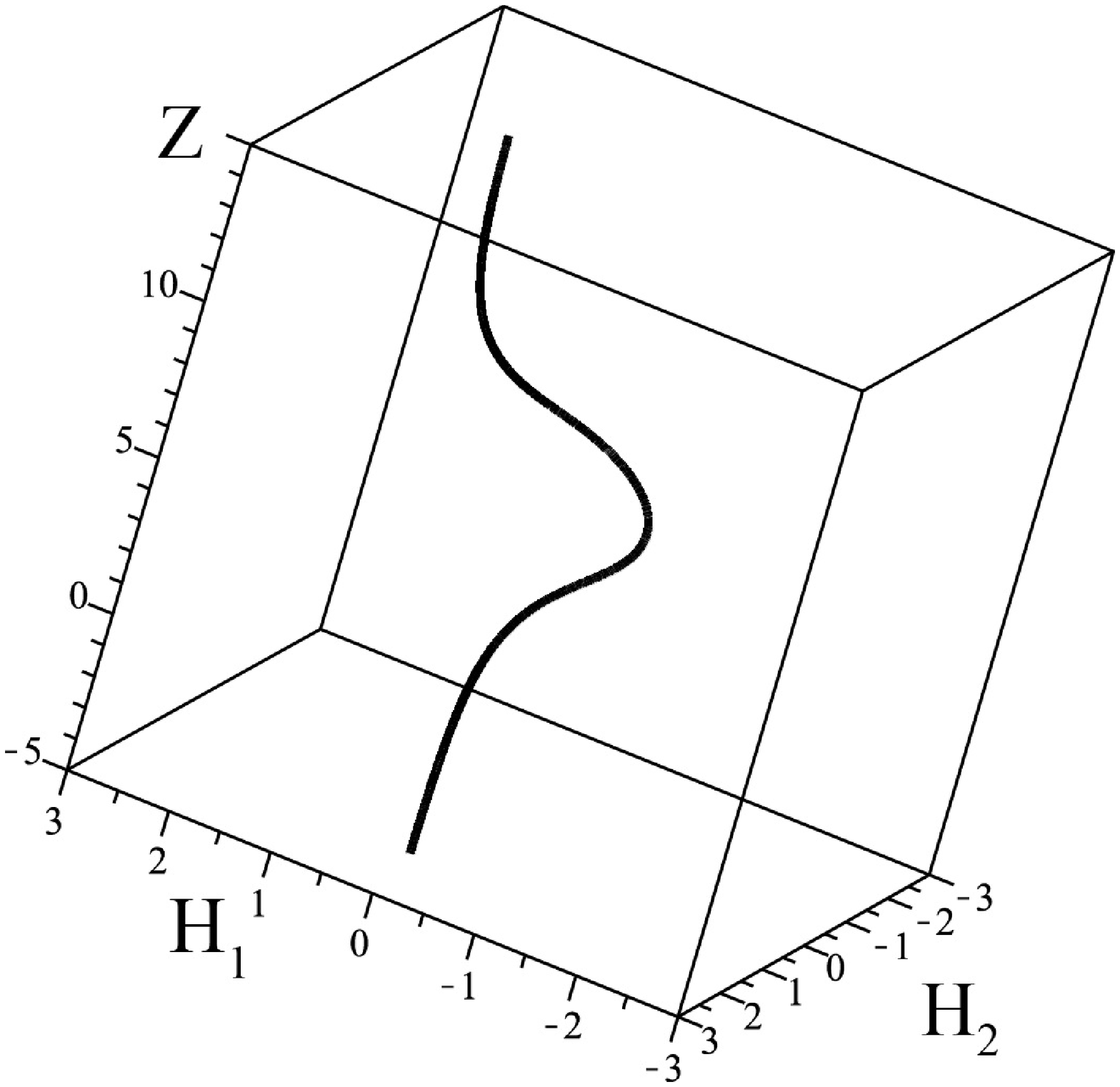}\\
  \caption{The numerical stationary solutions of equations (\ref{e48}) - (\ref{e49}) at $Ra \to 0$ in the form of nonlinear wave, soliton and kink. They correspond to the phase portraits, shown in Figure \ref{f4}.
}\label{f7}
\end{figure}

In Fig.\ref{f5} the corresponding area is presented with the number of fixed points. In the vertices of square area of the parameters $({C_3},{C_4})$ is located a fixed point of the elliptical type. The type of fixed point is determined by the form of phase portrait of Hamiltonian (\ref{e53}) for the constants ${C_3} =  \pm \chi$ and ${C_4} =  \pm \chi$ (see Fig.\ref{f6}). On the boundary of $({C_3},{C_4})$ are two fixed points: the elliptical and hyperbolic one. Phase portrait for these points is shown in Fig.\ref{f6}. For constants $({C_3} = 0,{C_4} = ([ - \chi ,0[,]0, - \chi ]))$, ${C_3} = ([ - \chi ,0[,]0, - \chi ]),{C_4} = 0)$ there are also two fixed points: the elliptical and hyperbolic one. The phase portrait for this point is shown in Fig.\ref{f6}. At last, within each sector of the area $({C_3},{C_4})$ are four fixed points, two elliptical and two hyperbolic. Phase portraits for these points, are shown in Fig.\ref{f6}. Fig.\ref{f7} shows a three-dimensional image of the limited stationary structures corresponding to the phase portraits presented in Fig.\ref{f7}. The left side of Fig.\ref{f7} shows the numerical solution corresponding to a nonlinear wave that occurs in the vicinity of elliptic point in the phase space. The central part of Fig.\ref{f7} shows the solitons solution, which correspond to the separatrix part going out and in the hyperbolic point. Finally, the right side of Fig.\ref{f7} presents a solution for the kink, which corresponds to the part of the separatrix linking two hyperbolic points.

Let us now turn to the evolution of large-scale stationary magnetic field for the case when the Rayleigh number $Ra \to 2$. In this limit, the Hamiltonian (\ref{e51}) takes the following form:
\[{\cal H} \to {\cal H}' = \frac{1}{{40Q}}\ln \left( {\frac{{{Q^2}H_1^4 + 4}}{{{Q^2}H_1^4 + 4QH_1^2 + 8}}} \right) + \frac{1}{{40Q}}\ln \left( {\frac{{{Q^2}H_2^4 + 4}}{{{Q^2}H_2^4 + 4QH_2^2 + 8}}} \right) + \]
\[ + \frac{1}{{10Q}}arctg\left( {\frac{{4(QH_1^2 + 1)}}{{4 - {Q^2}H_1^4 - 2QH_1^2}}} \right) + \frac{1}{{10Q}}arctg\left( {\frac{{4(QH_2^2 + 1)}}{{4 - {Q^2}H_2^4 - 2QH_2^2}}} \right)\]
\begin{equation}\label{e54}
   + {C_4}{H_1} - {C_3}{H_2} + {\widetilde C_5}
\end{equation}
At zero values of the constants ${C_3} = {C_4} = 0$ the phase portrait of Hamiltonian ((54)) is similar to the Fig.\ref{f4}. which implies the appearance of only one elliptic point in the phase space. In this case, all appearing stationary solutions coincide with the nonlinear waves. The stationary solution corresponding to the nonlinear finite amplitude wave are similar to the  Fig.\ref{f4}.

As in the previous case it is easy to set the parameters for the area ${C_3},{C_4}$ with a different number of fixed points. We  notice that the region of existence of fixed points of the parameter plane $({C_3},{C_4})$ is determined by the inequalities:
	\[ - \widetilde \chi  < {C_3} < \widetilde \chi \]
	\[ - \widetilde \chi  < {C_4} < \widetilde \chi \]
where $\widetilde \chi  = \max \left( {\frac{{{H_1}({Q^2}H_1^4 + RaQH_1^2 + 4)}}{{({Q^2}H_1^4 + 4)({Q^2}H_1^4 + 2RaQH_1^2 + R{a^2} + 4)}}} \right)$ when ${H_1}$is changing. For fixed values of the parameters $Ra,Q$, this maximum can be easily calculated. From the results of these calculations, we obtain the similar phase portraits of the Hamiltonian (54) and numerical solutions of equations (\ref{e48})-(\ref{e49}) at $Ra \to 2$ like in the previous case.
Thus, comparing the growth rate of the vortex and magnetic disturbances at the initial stage of large-scale development of the instability, we considered the emergence of large-scale stationary magnetic structures.  These structures are classified as stationary solutions of three types: nonlinear waves, solitons and kinks.

\section{Conclusions}
\label{si}

In this paper the closed system of nonlinear equations was obtained using the asymptotic method. These equations describe both linear and nonlinear increase stages of  hydrodynamic flows and magnetic fields in a electroconducting medium. This system of equations allows to explain emergence and stabilization of large-scale magnetic fields of some astrophysical objects, stars, in particular. It is also interesting to use it to describe the generation of large-scale fields by convection in electrically conductive medium in the interior of planets. It should be noted that despite the asymptotic technique based on the presence of small-scale oscillations  that is used, it is expected that the results can be applicable to the turbulent media. The turbulent case has the whole range of these small-scale oscillations. Qualitative evaluation of the linear stage for the solar conditions \cite{s25} allow us to state a good coincidence of the resulting hydrodynamic structures characteristic scales and times with observation data \cite{s38}. This fact suggests that other stationary magneto-hydrodynamic structures, like soliton exist in the Sun photosphere.

 \section*{Appendix I. Multiscale asymptotic developments}

Let us find the algebraic structure of the asymptotic development in various orders of $R$, starting with the lowest. In order of ${R^{ - 3}}$ there is only one equation:
\begin{equation}\label{e55}
  {\partial _i}{P_{ - 3}} = 0 \Rightarrow \;{P_{ - 3}} = {P_{ - 3}}(X)
\end{equation}
In order ${R^{ - 2}}$ appears equation:
\begin{equation}\label{e56}
  {\partial _i}{P_{ - 2}} = 0\quad \quad  \Rightarrow \;\;{P_{ - 2}} = {P_{ - 2}}\left( X \right)
\end{equation}
Consequently, quantities ${P_{ - 3}}$ and ${P_{ - 2}}$ depend only on fast variables. In order ${R^{ - 1}}$, we obtain more complicated system of equations:
\[{\partial _t}W_{ - 1}^i + W_{ - 1}^k{\partial _k}W_{ - 1}^i =  - {\partial _i}{P_{ - 1}} - {\nabla _i}{P_{ - 3}} + \partial _k^2W_{ - 1}^i +\]
\begin{equation}\label{e57}
   + \widetilde {Ra}{e_i}{T_{ - 1}} + \widetilde Q{\varepsilon _{ijk}}{\varepsilon _{jml}}{\partial _m}B_{ - 1}^lB_{ - 1}^k
\end{equation}
\begin{equation}\label{e58}
  {\partial _t}B_{ - 1}^i - P{m^{ - 1}}\partial _k^2B_{ - 1}^i = {\varepsilon _{ijk}}{\varepsilon _{knp}}{\partial _j}W_{ - 1}^nB_{ - 1}^p
\end{equation}
\begin{equation}\label{e59}
  {\partial _t}{T_{ - 1}} - P{r^{ - 1}}\partial _k^2{T_{ - 1}} =  - W_{ - 1}^k{\partial _k}{T_{ - 1}} - W_{ - 1}^z
\end{equation}
\begin{equation}\label{e60}
  {\partial _i}W_{ - 1}^i = 0,\quad \quad {\partial _i}B_{ - 1}^i = 0
\end{equation}
The averaging of equations (\ref{e57})-(\ref{e60}) on the fast variables give the following secular equations:
\begin{equation}\label{e61}
   - {\nabla _i}{P_{ - 3}} + \widetilde {Ra}{e_i}{T_{ - 1}} = 0
\end{equation}
\begin{equation}\label{e62}
  W_{ - 1}^z = 0
\end{equation}
In  zero order in $R$ we have the equations:
\[{\partial _t}v_0^i + W_{ - 1}^k{\partial _k}v_0^i + v_0^k{\partial _k}W_{ - 1}^i =  - {\partial _i}{P_0} - {\nabla _i}{P_{ - 2}} + \]
\begin{equation}\label{e63}
   + \partial _k^2v_0^i + \widetilde {Ra}{e_i}{T_0} + \widetilde Q{\varepsilon _{ijk}}{\varepsilon _{jml}}\left( {{\partial _m}B_{ - 1}^lB_0^k + {\partial _m}B_0^lB_{ - 1}^k} \right) + F_0^i
\end{equation}
\begin{equation}\label{e64}
  {\partial _t}B_0^i - P{m^{ - 1}}\partial _k^2B_0^i = {\varepsilon _{ijk}}{\varepsilon _{knp}}\left( {{\partial _j}W_{ - 1}^nB_0^p + {\partial _j}v_0^nB_{ - 1}^p} \right)
\end{equation}
\begin{equation}\label{e65}
  {\partial _t}{T_0} - P{r^{ - 1}}\partial _k^2{T_0} =  - W_{ - 1}^k{\partial _k}{T_0} - {\partial _k}(v_0^k{T_{ - 1}}) - v_0^z
\end{equation}
\begin{equation}\label{e66}
  {\partial _i}v_0^i = 0,\quad \quad {\partial _i}B_0^i = 0
\end{equation}
These equations give one secular equation:
\begin{equation}\label{e67}
  \nabla {P_{ - 2}} = 0\quad \quad  \Rightarrow \quad \quad {P_{ - 2}} = const
\end{equation}
Let us consider the equations of the first approximation ${R^1}$:
\[{\partial _t}v_1^i + W_{ - 1}^k{\partial _k}v_1^i + v_0^k{\partial _k}v_0^i + v_{1}^k{\partial _k}W_{ - {1}}^i + W_{ - {1}}^k{\nabla _k}W_{ - {1}}^i = \]
\[= - {\nabla _i}{P_{ - {1}}} - {\partial _i}\left( {{P_{1}} + {{\overline P }_{1}}} \right) + \partial _k^{2}{v_1}^i + 2{\partial _k}{\nabla _k}W_{ - 1}^i + \widetilde {Ra}{e_i} T_{1}+ \]
\begin{equation}\label{e68}
   + \widetilde Q{\varepsilon _{ijk}}{\varepsilon _{jml}}{\partial _m}B_{ - {1}}^lB_{1}^k + {\partial _m}B_0^lB_0^k + {\partial _m}B_{1}^lB_{ - {1}}^k + {\nabla _m}B_{ - {1}}^lB_{ - {1}}^k{)}
\end{equation}
\[{\partial _t}B_1^i - P{m^{ - 1}}\partial _k^2B_1^i - P{m^{ - 1}}2{\partial _k}{\nabla _k}B_{ - 1}^i = \]
\begin{equation}\label{e69}
   = {\varepsilon _{ijk}}{\varepsilon _{knp}}({\partial _j}W_{ - 1}^nB_1^p + {\partial _j}v_0^nB_0^p + {\partial _j}v_1^nB_{ - 1}^p + {\nabla _j}W_{ - 1}^nB_{ - 1}^p)
\end{equation}
\[{\partial _t}{T_1} - P{r^{ - 1}}\partial _k^2{T_1} - P{r^{ - 1}}2{\partial _k}{\nabla _k}{T_{ - 1}}=\]
\begin{equation}\label{e70}
   =  - W_{ - 1}^k{\partial _k}{T_1} - W_{ - 1}^k{\nabla _k}{T_{ - 1}} - v_0^k{\partial _k}{T_0} - v_1^k{\partial _k}{T_{ - 1}} - v_1^z
\end{equation}
\begin{equation}\label{e71}
  {\partial _i}v_1^i + {\nabla _i}W_{ - 1}^i = 0,\quad \quad {\partial _i}B_1^i + {\nabla _i}B_{ - 1}^i = 0
\end{equation}
The secular equations follow from this system of equations:
\begin{equation}\label{e72}
  W_{ - 1}^k{\nabla _k}W_{ - 1}^i =  - {\nabla _i}{P_{ - 1}} + \widetilde Q{\varepsilon _{ijk}}{\varepsilon _{jml}}{\nabla _m}B_{ - 1}^lB_{ - 1}^k
\end{equation}
\begin{equation}\label{e73}
  {\varepsilon _{ijk}}{\varepsilon _{knp}}{\nabla _j}W_{ - 1}^nB_{ - 1}^p = 0
\end{equation}
\begin{equation}\label{e74}
  W_{ - 1}^k{\nabla _k}{T_{ - 1}} = 0,
\end{equation}
\begin{equation}\label{e75}
  {\nabla _i}W_{ - 1}^i = 0,\quad \quad {\nabla _i}B_{ - 1}^i = 0
\end{equation}
The secular equation (\ref{e72})-(\ref{e75}) are satisfied by choosing the following geometry for the velocity and magnetic fields (Beltrami fields):
\[{\vec W_{ - 1}} = \left( {{W_x}(Z),{W_y}(Z),0} \right),\]
\begin{equation}\label{e76}
  {\vec B_{ - 1}} = \left( {B_{ - 1}^x(Z),B_{ - 1}^y(Z),0} \right),\quad {T_{ - 1}} = {T_{ - 1}}(Z),\quad {P_{ - 1}} = const
\end{equation}
In the second order ${R^2}$, we obtain the equations:
\[{\partial _t}v_{2}^i + W_{ - {1}}^k{\partial _k}v_{2}^i + v_0^k{\partial _k}v_{1}^i + W_{ - {1}}^k{\nabla _k}v_0^i + v_0^k{\nabla _k}W_{ - {1}}^i + v_{1}^k{\partial _k}v_0^i + v_{2}^k{\partial _k}W_{ - {1}}^i =   \]
\[= - {\nabla _i}{P_{2}} - {\nabla _i}{P_0} + \partial _k^2v_2^i + 2{\partial _k}{\nabla _k}v_0^i + \widetilde {Ra}{e_i}{T_2} + \widetilde Q{\varepsilon _{ijk}}{\varepsilon _{jml}}\left( {{\partial _m}B_{ - 1}^lB_2^k + } \right. \]
\begin{equation}\label{e77}
  + {\partial _m}B_0^lB_1^k + {\partial _m}B_1^lB_0^k + {\partial _m}B_2^lB_{ - 1}^k + \left. {{\nabla _m}B_{ - 1}^lB_0^k + {\nabla _m}B_0^lB_{ - 1}^k} \right)
\end{equation}
\[{\partial _t}B_2^i - P{m^{ - 1}}\partial _k^2B_2^i - P{m^{ - 1}}2{\partial _k}{\nabla _k}B_0^i = {\varepsilon _{ijk}}{\varepsilon _{knp}}\left( {{\partial _j}W_{ - {1}}^nB_{2}^p +  } \right.\]
\begin{equation}\label{e78}
  \left. { + {\partial _j}v_0^nB_{1}^p + {\partial _j}v_{1}^nB_0^p + {\partial _j}\;v_{2}^n\;B_{ - {1}}^p + {\nabla _j}W_{ - {1}}^nB_0^p + {\nabla _j}v_0^n\;B_{ - 1}^p} \right)
\end{equation}
\[{\partial _t}{T_2} - P{r^{ - 1}}\partial _k^2{T_2} - P{r^{ - 1}}2{\partial _k}{\nabla _k}{T_0} =  - W_{ - 1}^k{\partial _k}{T_2} - W_{ - 1}^k{\nabla _k}{T_0} -  \]
\begin{equation}\label{e79}
  - v_0^k{\partial _k}{T_1} - v_0^k{\nabla _k}{T_{ - 1}} - v_{1}^k{\partial _k}{T_0} - v_{2}^k{\partial _k}{T_{ - {1}}} - v_{2}^z
\end{equation}
\begin{equation}\label{e80}
  {\partial _i}v_2^i + {\nabla _i}v_0^i = 0,\quad \quad {\partial _i}B_2^i + {\nabla _i}B_0^i = 0
\end{equation}
It is easy to see that there are no secular terms in this order.
Let us consider now  the most important order ${R^3}$. In this order we obtain the equations:
\[{\partial _t}v_3^i + {\partial _T}W_{ - 1}^i + W_{ - 1}^k{\partial _k}v_3^i + v_0^k{\partial _k}v_2^i + \]
\[ + W_{ - 1}^k{\nabla _k}v_1^i + v_0^k{\nabla _k}v_0^i + v_1^k{\partial _k}v_1^i + v_2^k{\partial _k}v_0^i + v_1^k{\nabla _k}W_{ - 1}^i + v_3^k{\partial _k}W_{ - 1}^i = \]
\[ =  - {\partial _i}{P_{3}} - {\nabla _i}\left( {{P_{1}} + {{\overline P }_{1}}} \right) + \partial _k^{2}v_{3}^i + 2{\partial _k}{\nabla _k}v_{1}^i + \Delta W_{ - {1}}^i + \widetilde {Ra}{e_i}{T_{3}} + \]
\[+ \widetilde Q{\varepsilon _{ijk}}{\varepsilon _{jml}}\left( {{\partial _m}B_{ - {1}}^lB_{3}^k + {\partial _m}B_0^lB_{2}^k + {\partial _m}B_{1}^lB_{1}^k }\right. +\]
\begin{equation}\label{e81}
    + \left.{ {\partial _m}B_{2}^lB_0^k + {\nabla _m}B_{ - {1}}^lB_{1}^k + {\nabla _m}B_0^lB_0^k} \right) ,
\end{equation}
\[{\partial _t}B_{3}^i + {\partial _T}B_{ - {1}}^i - P{m^{ - {1}}}\partial _k^{2}B_{3}^i - P{m^{ - {1}}}{2}{\partial _k}{\nabla _k}B_{1}^i - P{m^{ - {1}}}\Delta B_{ - {1}}^i = \]
\[= {\varepsilon _{ijk}}{\varepsilon _{knp}}\left( {{\partial _j}W_{ - {1}}^nB_{3}^p + {\partial _j}v_0^nB_{2}^p + {\partial _j}v_{1}^nB_{1}^p + {\partial _j}v_{2}^nB_0^p +} \right.\]
\begin{equation}\label{e82}
   +\left. {{\nabla _j}W_{ - {1}}^nB_{1}^p + {\nabla _j}v_0^nB_0^p} \right),
\end{equation}
\[{\partial _t}{T_{3}} + {\partial _T}{T_{ - {1}}} - P{r^{ - {1}}}\partial _k^{2}{T_{3}} - P{r^{ - {1}}}{2}{\partial _k}{\nabla _k}{T_{1}} - P{r^{ - {1}}}\Delta {T_{ - {1}}} = \]
\[=  - W_{ - 1}^k{\partial _k}{T_3} - W_{ - 1}^k{\nabla _k}{T_1} - v_0^k{\partial _k}{T_2} - v_0^k{\nabla _k}{T_0} - v_1^k{\partial _k}{T_1}-\]
\begin{equation}\label{e83}
    - v_1^k{\nabla _k}{T_{ - 1}} - v_2^k{\partial _k}{T_0} - v_3^k{\partial _k}{T_{ - 1}} - v_3^z
\end{equation}
\begin{equation}\label{e84}
  {\partial _i}v_3^i + {\nabla _i}v_1^i = 0,\quad \quad {\partial _i}B_3^i + {\nabla _i}B_1^i = 0
\end{equation}
After averaging this system of equations over fast variables, we obtain the main system of secular equations to describe the evolution of large-scale perturbations:
\begin{equation}\label{e85}
  {\partial _t}W_{ - 1}^i - \Delta W_{ - 1}^i + {\nabla _k}\left( {\overline {v_0^kv_0^i} } \right) =  - {\nabla _i}{\overline P _1} + \widetilde Q{\varepsilon _{ijk}}{\varepsilon _{jml}}\left( {\overline {{\nabla _m}B_0^lB_0^k} } \right)
\end{equation}
\begin{equation}\label{e86}
  {\partial _t}B_{ - 1}^i - P{m^{ - 1}}\Delta B_{ - 1}^i = {\varepsilon _{ijk}}{\varepsilon _{knp}}{\nabla _j}\overline {(v_0^nB_0^p)}
\end{equation}
\begin{equation}\label{e87}
  {\partial _t}{T_{ - 1}} - P{r^{ - 1}}\Delta {T_{ - 1}} =  - {\nabla _k}(\overline {v_0^k{T_0}} )
\end{equation}
Using the well-known tensor identities:
\[{\varepsilon _{ijk}}{\varepsilon _{jml}} = {\delta _{km}}{\delta _{il}} - {\delta _{im}}{\delta _{kl}},\]
\[{\varepsilon _{ijk}}{\varepsilon _{knp}} = {\delta _{in}}{\delta _{jp}} - {\delta _{ip}}{\delta _{jp}}\]
and introducing for convenience designations $\vec W = {\vec W_{ - 1}}$, $\vec H = {\vec B_{ - 1}}$, $\Theta  = {T_{ - 1}}$, we write equations (\ref{e85})-(\ref{e87}) in the following form:
\begin{equation}\label{e88}
  {\partial _T}{W_i} - \Delta {W_i} + {\nabla _k}\overline {(v_0^kv_0^i)}  =  - {\nabla _i}\overline P  + \widetilde Q\left( {{\nabla _k}\left( {\overline {B_0^iB_0^k} } \right) - \frac{{{\nabla _i}}}{2}{{\left( {\overline {B_0^k} } \right)}^2}} \right)
\end{equation}
\begin{equation}\label{e89}
  {\partial _T}{H_i} - P{m^{ - {1}}}\Delta {H_i}{=}{\nabla _j}\left( {\overline {v_0^iB_0^j} } \right) - {\nabla _n}\left( {\overline {v_0^nB_0^i} } \right)
\end{equation}
\begin{equation}\label{e90}
  {\partial _T}\Theta  - P{r^{ - 1}}\Delta \Theta  + {\nabla _k}\left( {\overline {v_0^k{T_0}} } \right) = 0
\end{equation}

\section*{Appendix II. Small-scale fields}

In Appendix I, we obtain the equations in the zero order in $R$, which can be written using a more compact degnations for operators:
\[{\hat D_W}{=}{\partial _t} - {\partial ^{2}} + W_{ - 1}^k{\partial _k} ,\]
\begin{equation}\label{e91}
 {\hat D_H}{=}{\partial _t} - P{m^{ - {1}}}{\partial ^{2}} + W_{ - 1}^k{\partial _k}{,}\;{\hat D_\theta }{=}{\partial _t} - P{r^{ - {1}}}{\partial ^{2}} + W_{ - 1}^k{\partial _k},
\end{equation}
Then the set of equations (\ref{e63})-(\ref{e66}) takes the form:
\begin{equation}\label{e92}
  {\hat D_W}v_0^i =  - {\partial _i}{P_0} + \widetilde {Ra}{e_i}{T_0} + \widetilde Q{H_k}\left( {{\partial _k}B_0^i - {\partial _i}B_0^k} \right) + F_0^i,
\end{equation}
\begin{equation}\label{e93}
  {\hat D_H}B_0^i = {H_p}{\partial _p}v_0^i,
\end{equation}
\begin{equation}\label{e94}
  {\hat D_\theta }{T_0} =  - {e_k}v_0^k,
\end{equation}
\begin{equation}\label{e95}
  {\partial _i}v_0^i = {\partial _k}B_0^k = 0,
\end{equation}
It is not difficult to find expressions for the small-scale field ${\vec B_0}$ and ${T_0}$:
\begin{equation}\label{e96}
  B_0^i = \frac{{{H_p}{\partial _p}v_0^i}}{{{{\hat D}_H}}},\quad \quad {T_0} =  - \frac{{{e_k}v_0^k}}{{{{\hat D}_\theta }}}
\end{equation}
Now we substitute (\ref{e96}) in the equation (\ref{e92}), then differentiate the resulting expression for ${\partial _i}$, using the conditions of solenoidal fields (\ref{e95}). We obtain an expression for the pressure ${P_0}$:
\begin{equation}\label{e97}
  {P_0} =  - \frac{{\widetilde {Ra}{e_i}{e_k}{\partial _i}v_0^k}}{{{\partial ^2}{{\hat D}_\theta }}} - \frac{{\widetilde Q}}{{{\partial ^2}{{\hat D}_H}}}\left( {{H_p}{\partial _p}} \right)({H_k}{\partial ^2}v_0^k)
\end{equation}
Eliminating ${P_0}$ from (\ref{e92}) we obtain the equation for $v_0^k$:
\begin{equation}\label{e98}
  \left( {{\delta _{ik}} + \frac{{\widetilde {Ra}}}{{\hat q{{\hat D}_W}{{\hat D}_\theta }}}{{\hat P}_{ip}}{e_p}{e_k}} \right)v_0^k = \frac{{F_0^i}}{{\hat q{{\hat D}_W}}},
\end{equation}
where ${\hat P_{ip}}{=}{\delta _{ip}} - \frac{{{\partial _i}{\partial _p}}}{{{\partial ^{2}}}}$ is the projection operator,
\[\hat q{=}1 - \frac{{\widetilde Q{{{(}{H_k}{\partial _k}{\rm{)}}}^{2}}}}{{{{\hat D}_W}{{\hat D}_H}}} .\]
Let us rewrite (\ref{e98}) in the most compact form:
\begin{equation}\label{e99}
  {\hat L_{ik}}v_0^k = \frac{{F_0^i}}{{\hat q{{\hat D}_W}}},
\end{equation}
where the designation for the operator
\begin{equation}\label{e100}
  {\hat L_{ik}} = {\delta _{ik}} + \frac{{\widetilde {Ra}{{\hat P}_{ip}}{e_p}{e_k}}}{{\hat q{{\hat D}_W}{{\hat D}_\theta }}} .
\end{equation}
From (\ref{e99}) we can find  directly through the inverse operator $\hat L_{kj}^{ - 1}$ a velocity field $v_0^k$, i.e.
\begin{equation}\label{e101}
  v_0^k = L_{kj}^{ - 1}\frac{{F_0^j}}{{\hat q{{\hat D}_W}}},
\end{equation}
where $\hat L_{kj}^{ - 1}$ has the property ${\hat L_{ik}}\hat L_{kj}^{ - 1} = {\delta _{ij}}$:
\begin{equation}\label{e102}
  \hat L_{kj}^{ - 1} = {\delta _{kj}} - \frac{{\widetilde {Ra}{{\hat P}_{kn}}{e_n}{e_j}}}{{\hat q{{\hat D}_W}{{\hat D}_\theta } + \widetilde {Ra}{{\hat P}_{qp}}{e_q}{e_p}}}
\end{equation}
The expression for the small-scale fluctuations of velocity $v_0^k$ takes the form:
\begin{equation}\label{e103}
  v_0^k = \left[ {{\delta _{kj}} - \frac{{\widetilde {Ra}{{\hat P}_{kn}}{e_n}{e_j}}}{{\hat q{{\hat D}_W}{{\hat D}_\theta } + \widetilde {Ra}{{\hat P}_{qp}}{e_q}{e_p}}}} \right]\frac{{F_0^j}}{{\hat q{{\hat D}_W}}}
\end{equation}
Small-scale fluctuations of temperature ${T_0}$ expressed in terms of ${\vec v_0}({\vec F_0})$:
\begin{equation}\label{e104}
  {T_0} =  - \left[ {1 - \frac{{\widetilde {Ra}{{\hat P}_{kn}}{e_k}{e_n}}}{{\hat q{{\hat D}_W}{{\hat D}_\theta } + \widetilde {Ra}{{\hat P}_{qp}}{e_q}{e_p}}}} \right]\frac{{(\vec e{{\vec F}_0}{)}}}{{\hat q{{\hat D}_W}{{\hat D}_\theta }}}
\end{equation}
Equations (\ref{e103}) - (\ref{e104}) in the limit $Pr = 1$ and $\sigma  = 0$ (non-electroconductive medium) fully agree with the results of \cite{s18}, \cite{s19}. Next, we need to know the explicit form for small-scale fluctuations of magnetic fields ${\vec B_0}$:
\begin{equation}\label{e105}
  B_0^k = \left[ {{\delta _{kj}} - \frac{{\widetilde {Ra}{{\hat P}_{kn}}{e_n}{e_j}}}{{\hat q{{\hat D}_W}{{\hat D}_\theta } + \widetilde {Ra}{{\hat P}_{qp}}{e_q}{e_p}}}} \right]\frac{{{H_p}{\partial _p}F_0^j}}{{\hat q{D_W}{D_H}}}
\end{equation}
For further calculations of correlators we need to set explicitly the external helical force ${\vec F_0}$ in a deterministic form:
\begin{equation}\label{e106}
  {\vec F_0} = {f_0}\left[ {\vec i\cos {\varphi _2} + \vec j\sin {\varphi _1} + \vec k(\cos {\varphi _1} + \cos {\varphi _2})} \right]
\end{equation}
where ${\varphi _1} = {\vec \kappa _1}\vec x - {\omega _0}t$, ${\varphi _2} = {\vec \kappa _2}\vec x - {\omega _0}t$, ${\kappa _1} = {\kappa _0}\left( {1,0,0} \right),{\kappa _2} = {\kappa _0}(0,1,0)$.
Then the helicity of the external force is equal to:
\begin{equation}\label{e107}
  {\vec F_0}rot{\vec F_0} = {\kappa _0}\vec F_0^2 \ne 0
\end{equation}
It is convenient to rewrite the equation (\ref{e106}) in the complex form:
\begin{equation}\label{e108}
  {\vec F_0}{=}\vec A{e^{i{\varphi _{1}}}} + {\vec A^{*}}{e^{ - i{\varphi _{1}}}} + \vec B{e^{i{\varphi _{2}}}} + {\vec B^{*}}{e^{ - i{\varphi _{2}}}}
\end{equation}
where the vectors $\vec A$ and $\vec B$ are respectively:
\begin{equation}\label{e109}
  \vec A = \frac{{{f_0}}}{2}(\vec k - i\vec j),{\vec A^*} = \frac{{{f_0}}}{2}\left( {\vec k + i\vec j} \right),\vec B = \frac{{{f_0}}}{2}\left( {\vec i - i\vec k} \right),{\vec B^*}{=}\frac{{{f_0}}}{2}(\vec i + i\vec k)
\end{equation}
Action of the operators $\hat q$, ${\hat D_W}$, ${\hat D_H}$, ${\hat D_\theta }$ on their eigenfunctions $exp(i\omega t + i\vec \kappa \vec x)$  obviously has the form:
\[\hat q\left( {\omega ,\vec \kappa } \right)exp(i\vec \kappa \vec x + i\omega t),\quad {\hat D_W}\left( {\omega ,\vec \kappa } \right)\exp \left( {i\vec \kappa \vec x + i\omega t} \right)\;,\]
\[{\hat D_H}(\omega ,\vec \kappa )\exp \left( {i\vec \kappa \vec x + i\omega t} \right),\quad {\hat D_\theta }(\omega ,\vec \kappa )\exp \left( {i\vec \kappa \vec x + i\omega t} \right),\]
where $\hat q\left( {\omega ,\vec \kappa } \right){, }\;{\hat D_W}\left( {\omega ,\vec \kappa } \right){, }\;{\hat D_H}\left( {\omega ,\vec \kappa } \right)$, ${\hat D_\theta }{(}\omega ,\vec \kappa {)}$ have the form:
\begin{equation}\label{e110}
  \hat q\left( {\omega ,\vec \kappa } \right) = 1 + \frac{{\widetilde Q{{(\vec \kappa \vec H)}^2}}}{{{{\hat D}_W}\left( {\omega ,\vec \kappa } \right){{\hat D}_H}(\omega ,\vec \kappa )}}
\end{equation}
\[{\hat D_W}\left( {\omega ,\vec \kappa } \right) = i\left( {\omega  + \vec \kappa \vec W} \right) + {\kappa ^2}\]
\[{\hat D_H}\left( {\omega ,\vec \kappa } \right) = i\left( {\omega  + \vec \kappa \vec W} \right) + {\kappa ^{2}}P{m^{ - 1}}\]
\[{\hat D_\theta }\left( {\omega ,\vec \kappa } \right) = i\left( {\omega  + \vec \kappa \vec W} \right) + {\kappa ^2}P{r^{ - 1}}\]
We write down the number of useful relations, assuming for simplicity that ${\kappa _0} = 1$ and ${\omega _0} = 1$:
	\[{\hat D_W}\left( {{\omega _0}, - {{\vec \kappa }_{1}}} \right) = i\left( {1 - {W_1}} \right) + 1 = {\hat D_{{W_1}}}, {\hat D_W}\left( {{\omega _0}, - {{\vec \kappa }_1}} \right)= \hat D_{{W_1}}^*,\]
\begin{equation}\label{e111}
   {\hat D_H}\left( {{\omega _0}, - {{\vec \kappa }_{1}}} \right)= i\left( {1} - {W_{1}} \right) + P{m^{ - {1}}}={\hat D_{{H_{1}}}}, {\hat D_H}\left( {{\omega _0}, - {{\vec \kappa }_1}} \right) = \hat D_{{H_1}}^*,
\end{equation}
\[{\hat D_\theta }\left( {{\omega _0}, - {{\vec \kappa }_{1}}} \right) = i\left( {{1} - {W_{1}}} \right) + P{r^{ - 1}} = {\hat D_{{\theta _1}}},\quad {\hat D_\theta }\left( {{\omega _0}, - {{\vec \kappa }_{1}}} \right) = \hat D_{{\theta _1}}^*\]
\[\hat q\left( {{\omega _0}, - {{\vec \kappa }_1}} \right) = 1 + \frac{{\widetilde QH_1^2}}{{\left( {i\left( {1 - {W_1}} \right) + 1} \right)(i\left( {1 - {W_1}} \right) + P{m^{ - 1}})}} = {\hat q_1}\]
\[\hat q\left( { - {\omega _0},{{\vec \kappa }_1}} \right) = \hat q_1^*,\quad {H_1} = {H_x},\quad {W_1} = {W_x}\]
Similarly for the vector ${\vec \kappa _2}$ we get:
\[{\hat D_W}\left( {{\omega _0}, - {{\vec \kappa }_2}} \right) = i\left( {1 - {W_2}} \right) + 1 = {\hat D_{{W_2}}},\quad {\hat D_W}\left( {{\omega _0}, - {{\vec \kappa }_2}} \right) = \hat D_{{W_2}}^*,\]
\begin{equation}\label{e112}
  {\hat D_H}\left( {{\omega _0}{,} - {{\vec \kappa }_{2}}} \right) = i\left( {{1} - {W_{2}}} \right) + P{m^{ - 1}} = {\hat D_{{H_2}}},\quad {\hat D_H}\left( {{\omega _0}{,} - {{\vec \kappa }_{2}}} \right){=}\hat D_{{H_{2}}}^{*}
\end{equation}
\[{\hat D_\theta }\left( {{\omega _0}, - {{\vec \kappa }_2}} \right) = i\left( {1 - {W_2}} \right) + P{r^{ - 1}} = {\hat D_{{\theta _2}}},\quad {\hat D_\theta }\left( {{\omega _0}, - {{\vec \kappa }_2}} \right) = \hat D_{{\theta _2}}^*\]
\[\hat q\left( {{\omega _0}, - {{\vec \kappa }_2}} \right) = 1 + \frac{{\widetilde QH_2^2}}{{\left( {i\left( {1 - {W_2}} \right) + 1} \right)(i\left( {1 - {W_2}} \right) + P{m^{ - 1}})}} = {\hat q_2}\]
\[\hat q\left( { - {\omega _0},{{\vec \kappa }_2}} \right) = \hat q_2^*,\quad {H_2} = {H_y},{W_2} = {W_y}\]
According to the definition of external force ${\vec F_0}$ (\ref{e108}), the small-scale fields ${\vec v_0}$, ${\vec B_0}$, ${T_0}$ determined by formulas (\ref{e103})-(\ref{e105}), each of them consist of four terms:
\[v_0^k{=}v_{{0}1}^k + v_{{0}2}^k + v_{{0}3}^k + v_{{0}4}^k;\]
\begin{equation}\label{e113}
  B_0^i = B_{01}^i + B_{02}^i + B_{03}^i + B_{04}^i;\quad {T_0} = {T_{01}} + {T_{02}} + {T_{03}} + {T_{04}};
\end{equation}
where $v_{{0}2}^k = (v_{01}^k{)^*}$, $v_{04}^k = (v_{03}^k{)^*}$, $B_{02}^k = (B_{01}^k{)^*}$, $B_{04}^k = (B_{03}^k{)^*}$, ${T_{02}} = ({T_{01}}{)^*}$, ${T_{04}} = ({T_{03}}{)^*}$
\begin{equation}\label{e114}
  v_{01}^k = {e^{i{\varphi _1}}}\left[ {{\delta _{kj}} - \frac{{\widetilde {Ra}{{\hat P}_{kn}}{e_n}{e_j}}}{{\hat q_1^*\hat D_{{W_1}}^*\hat D_{{\theta _1}}^* + \widetilde {Ra}}}} \right]\frac{{{A_j}}}{{\hat q_1^*\hat D_{{W_1}}^*}}
\end{equation}
\begin{equation}\label{e115}
  v_{{0}3}^k{=}{e^{i{\varphi _2}}}\left[ {{\delta _{kj}} - \frac{{\widetilde {Ra}{{\hat P}_{kn}}{e_n}{e_j}}}{{\hat q_2^*\hat D_{{W_2}}^*\hat D_{{\theta _2}}^* + \widetilde {Ra}}}} \right]\frac{{{B_j}}}{{\hat q_{2}^*\hat D_{{W_2}}^*}}
\end{equation}
\begin{equation}\label{e116}
  {T_{01}} =  - {e^{i{\varphi _1}}}\left[ {1 - \frac{{\widetilde {Ra}{{\hat P}_{kn}}{e_n}{e_k}}}{{\hat q_1^*\hat D_{{W_1}}^*\hat D_{{\theta _1}}^* + \widetilde {Ra}}}} \right]\frac{{(\vec e\vec A)}}{{\hat q_1^*\hat D_{{W_1}}^*\hat D_{{\theta _1}}^*}}
\end{equation}
\begin{equation}\label{e117}
  {T_{{0}3}} =  - {e^{i{\varphi _2}}}\left[ {1 - \frac{{\widetilde {Ra}{{\hat P}_{kn}}{e_n}{e_k}}}{{\hat q_2^*\hat D_{{W_2}}^*\hat D_{{\theta _2}}^* + \widetilde {Ra}}}} \right]\frac{{(\vec e\vec B)}}{{\hat q_{2}^*\hat D_{{W_2}}^*\hat D_{{\theta _2}}^*}}
\end{equation}
\begin{equation}\label{e118}
  B_{01}^k = {e^{i{\varphi _1}}}\left[ {{\delta _{kj}} - \frac{{\widetilde {Ra}{{\hat P}_{kn}}{e_n}{e_j}}}{{\hat q_1^*\hat D_{{W_1}}^*\hat D_{{\theta _1}}^* + \widetilde {Ra}}}} \right]\frac{{i{H_1}{A_j}}}{{\hat q_1^*\hat D_{{W_1}}^*\hat D_{{H_1}}^*}}
\end{equation}
\begin{equation}\label{e119}
  B_{03}^k = {e^{i{\varphi _2}}}\left[ {{\delta _{kj}} - \frac{{\widetilde {Ra}{{\hat P}_{kn}}{e_n}{e_j}}}{{\hat q_2^*\hat D_{{W_2}}^*\hat D_{{\theta _2}}^* + \widetilde {Ra}}}} \right]\frac{{i{H_2}{B_j}}}{{\hat q_2^*\hat D_{{W_2}}^*\hat D_{{H_2}}^*}}
\end{equation}
 Here we take into account that ${\hat P_{qs}}\left( {{\kappa _1}} \right){e_q}{e_s} = {\hat P_{qs}}\left( {{\kappa _2}} \right){e_q}{e_s} = 1$ because ${\vec \kappa _1},{\vec \kappa _2} \bot \vec e$.

\section*{Appendix III. Reynolds stress, Maxwell stress and turbulent e.m.f.}

Let us start with the calculation of Reynolds stresses $\overline {v_0^kv_0^i}  = T_{(1)}^{ki} + T_{(2)}^{ki}$. We need the equations (\ref{e114})-(\ref{e115}) and type of the external helical  force (\ref{e108})-(\ref{e109}). For simplicity, we assume that the dimensionless amplitude of the external helical force ${f_0} = 1$. Then we have:
\[T_{(1)}^{ki} = \frac{1}{{{{\left| {{{\hat q}_1}} \right|}^2}{{\left| {{{\hat D}_{{W_1}}}} \right|}^2}}}\left\{ {\left( {{A_k}A_i^* + {A_i}A_k^*} \right) - \hat aA_z^*\left( {{e_i}{A_k} + {e_k}{A_i}} \right) - } \right.\]
\begin{equation}\label{e120}
  \left. { - {{\hat a}^*}{A_z}\left( {{e_i}A_k^* + {e_k}A_i^*} \right) + 2{{\left| {\hat a} \right|}^2}{{\left| {{A_z}} \right|}^2}{e_i}{e_k}} \right\},
\end{equation}
Where are introduced the following designations:
\[{\left| {{{\hat q}_1}} \right|^2} = {\hat q_1}\hat q_1^* = \hat q_1^*{\hat q_1} = \frac{{{{(P{m^{ - 1}} - {{\left( {1 - {W_1}} \right)}^2} + \widetilde QH_1^2)}^2} + \left( {1 - {W_1}} \right){{(1 + P{m^{ - 1}})}^2}}}{{\left( {1 + {{\left( {1 - {W_1}} \right)}^2}} \right)(P{m^{ - 2}} + {{\left( {1 - {W_1}} \right)}^2})}};\]
\begin{equation}\label{e121}
  {\left| {{{\hat D}_{{W_1}}}} \right|^2} = \hat D_{{W_1}}^*{\hat D_{{W_1}}} = 1 + {(1 - {W_1})^2}
\end{equation}
\[{\hat a^*} = \frac{{\widetilde {Ra}}}{{\hat q_1^*\hat D_{{W_1}}^*\hat D_{{\theta _1}}^* + \widetilde {Ra}}},\hat a = \frac{{\widetilde {Ra}}}{{{{\hat q}_1}{{\hat D}_{{W_1}}}{{\hat D}_{{\theta _1}}} + \widetilde {Ra}}},{\left| {\hat a} \right|^2} = \hat a{\hat a^*},{\left| {{A_z}} \right|^2} = A_z^*{A_z}\]
The expression for the $T_{(2)}^{ki}$ has the similar form:
\begin{equation}\label{e122}
  T_{(2)}^{ki} = \frac{1}{{{{\left| {{{\hat q}_2}} \right|}^2}{{\left| {{{\hat D}_{{W_2}}}} \right|}^2}}}\left\{ {\left( {{B_k}B_i^* + {B_i}B_k^*} \right) - \hat bB_z^*\left( {{e_i}{B_k} + {e_k}{B_i}} \right) - } \right.
\end{equation}
\[\left. { - {{\hat b}^*}{B_z}\left( {{e_i}B_k^* + {e_k}B_i^*} \right) + 2{{\left| {\hat b} \right|}^2}{{\left| {{B_z}} \right|}^2}{e_i}{e_k}} \right\},\]
Here
\[{\left| {{{\hat q}_2}} \right|^2} = {\hat q_2}\hat q_2^* = \hat q_2^*{\hat q_2} =\]
\begin{equation}\label{e123}
  = \frac{{{{(P{m^{ - 1}} - {{\left( {1 - {W_2}} \right)}^2} + \widetilde QH_2^2)}^2} + \left( {1 - {W_2}} \right){{(1 + P{m^{ - 1}}{)}}^2}}}{{\left( {1 + {{\left( {1 - {W_2}} \right)}^2}} \right)(P{m^{ - 2}} + {{\left( {1 - {W_2}} \right)}^2})}}
\end{equation}
\[{\left| {{{\hat D}_{{W_2}}}} \right|^2} = \hat D_{{W_2}}^*{\hat D_{{W_2}}} = 1 + {(1 - {W_2})^2};\]
\begin{equation}\label{e124}
  {\hat b^*} = \frac{{\widetilde {Ra}}}{{\hat q_2^*\hat D_{{W_2}}^*\hat D_{{\theta _2}}^* + \widetilde {Ra}}},\hat b = \frac{{\widetilde {Ra}}}{{{{\hat q}_2}{{\hat D}_{{W_2}}}{{\hat D}_{{\theta _2}}} + \widetilde {Ra}}},{\left| {\hat b} \right|^2} = \hat b{\hat b^*},{\left| {{B_z}} \right|^2} = B_z^*{B_z}
\end{equation}
Equations (\ref{e120})-(\ref{e124}), in the limit of non electroconductive medium ($\sigma = 0$) and $Pr = 1$, were obtained in \cite{s18}, \cite{s19}.
Correlators for magnetic fields (Maxwell stress) $S_{(1)}^{ik} + S_{(2)}^{ik}$ can be found using (\ref{e118})-(\ref{e119}):
\[S_{(1)}^{ki} = \frac{{H_1^2}}{{{{\left| {{{\hat q}_1}} \right|}^2}{{\left| {{{\hat D}_{{W_1}}}} \right|}^2}{{\left| {{{\hat D}_{{H_1}}}} \right|}^2}}}\left\{ {\left( {{A_i}A_k^* + {A_k}A_i^*} \right) - \hat aA_z^*\left( {{e_k}{A_i} + {e_i}{A_k}} \right) -  } \right.\]
\begin{equation}\label{e125}
  \left. {- {{\hat a}^*}{A_z}\left( {{e_i}A_k^* + {e_k}A_i^*} \right) + 2{{\left| {\hat a} \right|}^2}{{\left| {{A_z}} \right|}^2}{e_i}{e_k}} \right\}
\end{equation}
 where ${\left| {{{\hat D}_{{H_1}}}} \right|^2}{=}\hat D_{{H_1}}^*{\hat D_{{H_1}}} = P{m^{ - 2}} + {(1 - {W_1})^2}$;
\[S_{(2)}^{ki} = \frac{{H_2^2}}{{{{\left| {{{\hat q}_2}} \right|}^2}{{\left| {{{\hat D}_{{W_2}}}} \right|}^2}{{\left| {{{\hat D}_{{H_2}}}} \right|}^2}}}\left\{ {\left( {{B_i}B_k^* + {B_k}B_i^*} \right) - \hat bB_z^*\left( {{e_k}{B_i} + {e_i}{B_k}} \right) - } \right.\]
\begin{equation}\label{e126}
  \left. { - {{\hat b}^*}{B_z}\left( {{e_i}B_k^* + {e_k}B_i^*} \right) + 2{{\left| {\hat b} \right|}^2}{{\left| {{B_z}} \right|}^2}{e_i}{e_k}} \right\}
\end{equation}
 where ${\left| {{{\hat D}_{{H_2}}}} \right|^2} = \hat D_{{H_2}}^*{\hat D_{{H_2}}} = P{m^{ - 2}} + {(1 - {W_2})^2}$. Using (114)-(115) and (118)-(119), we obtain expressions for the mixed correlator $G_{(1)}^{ik} + G_{(2)}^{ik}$:
\[G_{(1)}^{ik} = \frac{{iP{m^{ - 1}}{H_1}}}{{{{\left| {{{\hat q}_1}} \right|}^2}{{\left| {{{\hat D}_{{W_1}}}} \right|}^2}{{\left| {{{\hat D}_{{H_1}}}} \right|}^2}}}\times\]
\[\times\left\{ {\left( {{A_j}A_i^* - {A_i}A_j^*} \right) - \hat aA_z^*\left( {{e_i}{A_j} - {e_j}{A_i}} \right) - {{\hat a}^*}{A_z}\left( {{e_j}A_i^* - {e_i}A_j^*} \right)} \right\} - \]
\[ - \frac{{(1 - {W_1}){H_1}}}{{{{\left| {{{\hat q}_1}} \right|}^2}{{\left| {{{\hat D}_{{W_1}}}} \right|}^2}{{\left| {{{\hat D}_{{H_1}}}} \right|}^2}}}\left\{ {\left( {{A_j}A_i^* + {A_i}A_j^*} \right) - \hat aA_z^*\left( {{e_i}{A_j} + {e_j}{A_i}} \right) -  } \right.\]
\begin{equation}\label{e127}
  \left. {- {{\hat a}^*}{A_z}\left( {{e_j}A_i^* + {e_i}A_j^*} \right) + 2{{\left| {\hat a} \right|}^2}{{\left| {{A_z}} \right|}^2}{e_i}{e_j}} \right\}
\end{equation}
\[G_{(2)}^{ij} = \frac{{iP{m^{ - 1}}{H_2}}}{{{{\left| {{{\hat q}_2}} \right|}^2}{{\left| {{{\hat D}_{{W_2}}}} \right|}^2}{{\left| {{{\hat D}_{{H_2}}}} \right|}^2}}} \times\]
\[\times\left\{ {\left( {{B_j}B_i^* - {B_i}B_j^*} \right) - \hat bB_z^*\left( {{e_i}{B_j} - {e_j}{B_i}} \right) - {{\hat b}^*}{B_z}\left( {{e_j}B_i^* - {e_i}B_j^*} \right)} \right\} - \]
\[ - \frac{{(1 - {W_2}){H_2}}}{{{{\left| {{{\hat q}_2}} \right|}^2}{{\left| {{{\hat D}_{{W_2}}}} \right|}^2}{{\left| {{{\hat D}_{{H_2}}}} \right|}^2}}}\left\{ {\left( {{B_j}B_i^* + {B_i}B_j^*} \right) - \hat bB_z^*\left( {{e_i}{B_j} + {e_j}{B_i}} \right) -  } \right.\]
\begin{equation}\label{e128}
  \left. {- {{\hat b}^*}{B_z}\left( {{e_j}B_i^* + {e_i}B_j^*} \right) + 2{{\left| {\hat b} \right|}^2}{{\left| {{B_z}} \right|}^2}{e_i}{e_j}} \right\}
\end{equation}
By simple replacement of indices $i \to n$, $j \to i$ we obtain the expressions for $G_{(1)}^{ni}$ and $G_{(2)}^{ni}$. Since we are interested in the evolution of large-scale fields $\vec W$ and $\vec H$, and taking into account the geometry of the problem (\ref{e20}), we need to know the following Reynolds stress components:
\[T_{(1)}^{31} + T_{(2)}^{31}; \; T_{(1)}^{32} + T_{(2)}^{32}; \; S_{(1)}^{31} + S_{(2)}^{31}; \; S_{(1)}^{32} + S_{(2)}^{32}; \]
\begin{equation}\label{e129}
  G_{(1)}^{13} + G_{(2)}^{13}; \; G_{(1)}^{31} + G_{(2)}^{31}; \; G_{(1)}^{23} + G_{(2)}^{23}; \; G_{(1)}^{32} + G_{(2)}^{32}
\end{equation}
Using the equation (\ref{e120}) we can find the  expression for $T_{(1)}^{31}$, while putting $k = 3$, $i = 1$:
\[T_{(1)}^{31} = \frac{1}{{{{\left| {{{\hat q}_1}} \right|}^2}{{\left| {{{\hat D}_{{W_1}}}} \right|}^2}}}\left\{ {\left( {{A_3}A_1^* + {A_1}A_3^*} \right) - \hat aA_z^*\left( {{e_1}{A_3} + {e_3}{A_1}} \right) - } \right.\]
\begin{equation}\label{e130}
  \left. { - {{\hat a}^*}{A_z}\left( {{e_1}A_3^* + {e_3}A_1^*} \right) + 2{{\left| {\hat a} \right|}^2}{{\left| {{A_z}} \right|}^2}{e_1}{e_3}} \right\}
\end{equation}
 since ${e_1} = 0$ and ${A_1} = A_1^* = 0$, ${A_z} = {A_3}$, $T_{(1)}^{31} = 0$.
In a similar way is calculated $T_{(2)}^{31}$ using (\ref{e122}), where indices $k = 3$ and $i = 1$, i.e.
\[T_{(2)}^{31} = \frac{1}{{{{\left| {{{\hat q}_2}} \right|}^2}{{\left| {{{\hat D}_{{W_2}}}} \right|}^2}}}\left\{ {\left( {{B_3}B_1^* + {B_1}B_3^*} \right) - \hat bB_z^*\left( {{e_1}{B_3} + {e_3}{B_1}} \right) - } \right.\]
\[\left. { - {{\hat b}^*}{B_z}\left( {{e_1}B_3^* + {e_3}B_1^*} \right) + 2{{\left| {\hat b} \right|}^2}{{\left| {{B_z}} \right|}^2}{e_1}{e_3}} \right\}\]
\begin{equation}\label{e131}
  {B_z} = {B_3}
\end{equation}
Considering, that ${B_3}B_1^* + {B_1}B_3^* = 0$ and ${e_1} = 0$, the equation (\ref{e131}) is simplified:
\begin{equation}\label{e132}
  T_{(2)}^{31} = \frac{i}{{4{{\left| {{{\hat q}_2}} \right|}^2}{{\left| {{{\hat D}_{{W_2}}}} \right|}^2}}}\left( {{{\hat b}^*} - \hat b} \right)
\end{equation}
or after substitution of (\ref{e124}), we have:
\begin{equation}\label{e133}
  T_{(2)}^{31} =  - \frac{i}{4}\frac{{\widetilde {Ra}}}{{{{\left| {{{\hat q}_2}} \right|}^2}{{\left| {{{\hat D}_{{W_2}}}} \right|}^2}}}\left\{ {\frac{{\hat q_2^*\hat D_{{W_2}}^*\hat D_{{\theta _2}}^* - {{\hat q}_2}{{\hat D}_{{W_2}}}{{\hat D}_{{\theta _2}}}}}{{{{\left| {{{\hat q}_2}{{\hat D}_{{W_2}}}{{\hat D}_{{\theta _2}}} + \widetilde {Ra}} \right|}^2}}}} \right\}
\end{equation}
Taking in formulas (\ref{e120}) and (\ref{e122}) indices $k$ and $i$ equal respectively $k = 3$ and $i = 2$, we can obtain the expression for the $T_{(1)}^{31}$ and $T_{(2)}^{31}$:
\[T_{(1)}^{32} = \frac{1}{{{{\left| {{{\hat q}_1}} \right|}^2}{{\left| {{{\hat D}_{{W_1}}}} \right|}^2}}}\left\{ {\left( {{A_3}A_2^* + {A_2}A_3^*} \right) - \hat aA_z^*\left( {{e_2}{A_3} + {e_3}{A_2}} \right) - } \right.\]
\begin{equation}\label{e134}
  \left. { - {{\hat a}^*}{A_z}\left( {{e_2}A_3^* + {e_3}A_2^*} \right) + 2{{\left| {\hat a} \right|}^2}{{\left| {{A_z}} \right|}^2}{e_2}{e_3}} \right\}
\end{equation}
given the  ${e_2} = 0$, ${A_3}A_2^* + {A_2}A_3^* = 0$ and equation (\ref{e121}), the form of the expression (\ref{e134}) becomes simpler:
\begin{equation}\label{e135}
  T_{(1)}^{{3}2} = \frac{i}{4}\frac{{\widetilde {Ra}}}{{{{\left| {{{\hat q}_1}} \right|}^2}{{\left| {{{\hat D}_{{W_1}}}} \right|}^2}}}\left\{ {\frac{{\hat q_1^*\hat D_{{W_1}}^*\hat D_{{\theta _1}}^* - {{\hat q}_1}{{\hat D}_{{W_1}}}{{\hat D}_{{\theta _1}}}}}{{{{\left| {{{\hat q}_1}{{\hat D}_{{W_1}}}{{\hat D}_{{\theta _1}}} + \widetilde {Ra}} \right|}^2}}}} \right\}
\end{equation}
Component $T_{(2)}^{32}$ is zero due to the fact that the ${e_2} = 0$ and ${B_2} = B_2^* = 0$:
\[T_{(2)}^{32} = \frac{1}{{{{\left| {{{\hat q}_2}} \right|}^2}{{\left| {{{\hat D}_{{W_2}}}} \right|}^2}}}\left\{ {\left( {{B_3}B_2^* + {B_2}B_3^*} \right) - \hat bB_z^*\left( {{e_2}{B_3} + {e_3}{B_2}} \right) - } \right.\]
\begin{equation}\label{e136}
  \left. { - {{\hat b}^*}{B_z}\left( {{e_2}B_3^* + {e_3}B_2^*} \right) + 2{{\left| {\hat b} \right|}^2}{{\left| {{B_z}} \right|}^2}{e_2}{e_3}} \right\}
\end{equation}
In the equations (\ref{e125}) and (\ref{e126}) we take indices $k$ and $i$ equal to $k = 3$, $i = 1$.  Then the expressions can be found for the components  $S_{(1)}^{31}$ and $S_{(2)}^{31}$, respectively:
\[S_{\left( 1 \right)}^{31} = \frac{{H_1^2}}{{{{\left| {{{\hat q}_1}} \right|}^2}{{\left| {{{\hat D}_{{W_1}}}} \right|}^2}{{\left| {{{\hat D}_{{H_1}}}} \right|}^2}}}\left\{ {\left( {{A_3}A_1^* + {A_1}A_3^*} \right) - \hat aA_z^*\left( {{e_k}{A_i} + {e_i}{A_k}} \right) - } \right.\]
\begin{equation}\label{e137}
  \left. {-{{\hat a}^*}{A_3}\left( {{e_1}A_3^* + {e_3}A_1^*} \right) + 2{{\left| {\hat a} \right|}^2}{{\left| {{A_3}} \right|}^2}{e_1}{e_3}} \right\} = 0
\end{equation}
i.e. ${e_1} = 0$ and ${A_1} = A_1^* = 0;$
\[S_{(2)}^{31} = \frac{{H_2^2}}{{{{\left| {{{\hat q}_2}} \right|}^2}{{\left| {{{\hat D}_{{W_2}}}} \right|}^2}{{\left| {{{\hat D}_{{H_2}}}} \right|}^2}}}\left\{ {\left( {{B_3}B_1^* + {B_1}B_3^*} \right) - \hat bB_3^*\left( {{e_1}{B_3} + {e_3}{B_1}} \right) -  } \right.\]
\begin{equation}\label{e138}
  \left. { - {{\hat b}^*}{B_3}\left( {{e_3}B_1^* + {e_1}B_3^*} \right) + 2{{\left| {\hat b} \right|}^2}{{\left| {{B_z}} \right|}^2}{e_3}{e_1}} \right\}
\end{equation}
We take into account that ${B_3}B_1^* + {B_1}B_3^* = 0$ and ${e_1} = 0$, then (\ref{e138}) takes the form:
\begin{equation}\label{e139}
  S_{(2)}^{31} = \frac{i}{4}\frac{{H_2^2\widetilde {Ra}}}{{{{\left| {{{\hat q}_2}} \right|}^2}{{\left| {{{\hat D}_{{H_2}}}} \right|}^2}{{\left| {{{\hat D}_{{W_2}}}} \right|}^2}}}\left\{ {\frac{{{{\hat q}_2}{{\hat D}_{{W_2}}}{{\hat D}_{{\theta _2}}} - \hat q_2^*\hat D_{{W_2}}^*\hat D_{{\theta _2}}^*}}{{{{\left| {{{\hat q}_2}{{\hat D}_{{W_2}}}{{\hat D}_{{\theta _2}}} + \widetilde {Ra}} \right|}^2}}}} \right\}
\end{equation}
From (\ref{e125}) and (\ref{e126}) we have the  equations for the components of $S_{(1)}^{32}$ and $S_{(2)}^{32}$ , assuming that $k = 3$, $i =2$. Then
\[S_{(1)}^{32} = \frac{{H_1^2}}{{{{\left| {{{\hat q}_{1}}} \right|}^2}{{\left| {{{\hat D}_{{W_{1}}}}} \right|}^2}{{\left| {{{\hat D}_{{H_1}}}} \right|}^2}}}\left\{ {\left( {{A_3}A_2^* + {A_2}A_3^*} \right) - \hat aA_3^*\left( {{e_2}{A_3} + {e_3}{A_2}} \right) - } \right.\]
\[\left. {-{{\hat a}^*}{A_3}\left( {{e_3}A_2^* + {e_2}A_3^*} \right) +  2{{\left| {\hat a} \right|}^2}{{\left| {{A_3}} \right|}^2}{e_3}{e_2}} \right\}\]
here ${A_3}A_2^* + {A_2}A_3^* = 0$, ${e_2} = 0$. Then $S_{(1)}^{32}$ has the form:
\begin{equation}\label{e140}
  S_{{(}1)}^{32} = \frac{i}{4}\frac{{H_1^2\widetilde {Ra}}}{{{{\left| {{{\hat q}_1}} \right|}^2}{{\left| {{{\hat D}_{{W_1}}}} \right|}^2}{{\left| {{{\widehat {D}}_{{H_1}}}} \right|}^2}}}\left\{ {\frac{{\hat q_1^*\hat D_{{W_1}}^*\hat D_{{\theta _1}}^* - {{\hat q}_1}{{\hat D}_{{W_1}}}{{\hat D}_{{\theta _1}}}}}{{{{\left| {{{\hat q}_1}{{\hat D}_{{W_1}}}{{\hat D}_{{\theta _1}}} + \widetilde {Ra}} \right|}^2}}}} \right\}
\end{equation}
Component $S_{(2)}^{32}$ is equal to zero, i.e. ${E_2} = 0$ and ${B_2} = B_2^* = 0$:
\[S_{(2)}^{32} = \frac{{H_2^2}}{{{{\left| {{{\hat q}_2}} \right|}^2}{{\left| {{{\hat D}_{{W_2}}}} \right|}^2}{{\left| {{{\hat D}_{{H_2}}}} \right|}^2}}}\left\{ {\left( {{B_3}B_2^* + {B_2}B_3^*} \right) - \hat bB_3^*\left( {{e_2}{B_3} + {e_3}{B_1}} \right) -  } \right.\]
\begin{equation}\label{e141}
  \left. { -{{\hat b}^*}{B_3}\left( {{e_3}B_2^* + {e_2}B_3^*} \right) + 2{{\left| {\hat b} \right|}^2}{{\left| {{B_3}} \right|}^2}{e_3}{e_2}} \right\} = 0
\end{equation}
Further, according to the equations (\ref{e127}) and (\ref{e128}) and  replacing indices $i$ and $j$ by $i = 1$ and $j = 3$ ; $i = 3$ and $j = 1$ ; $i = 2$ and $j = 3$ ; $i = 3$ and $j = 2$, respectively, we get:
\[G_{(1)}^{13} = \frac{{iP{m^{ - 1}}{H_1}}}{{{{\left| {{{\hat q}_1}} \right|}^2}{{\left| {{{\hat D}_{{W_1}}}} \right|}^2}{{\left| {{{\hat D}_{{H_1}}}} \right|}^2}}} \times\]
\[\times \left\{ {\left( {{A_3}A_1^* - {A_1}A_3^*} \right) - \hat aA_3^*\left( {{e_1}{A_3} - {e_3}{A_1}} \right) - {{\hat a}^*}{A_3}\left( {{e_3}A_1^* - {e_1}A_3^*} \right)} \right\} - \]
\[ - \frac{{(1 - {W_1}{)}{H_1}}}{{{{\left| {{{\hat q}_1}} \right|}^2}{{\left| {{{\hat D}_{{W_1}}}} \right|}^2}{{\left| {{{\hat D}_{{H_1}}}} \right|}^2}}}\left\{ {\left( {{A_3}A_1^* + {A_1}A_3^*} \right) - \hat aA_3^*\left( {{e_1}{A_3} + {e_3}{A_1}} \right) - } \right.\]
\begin{equation}\label{e142}
  \left. {-  {{\hat a}^*}{A_3}\left( {{e_3}A_1^* + {e_1}A_3^*} \right) + 2{{\left| {\hat a} \right|}^2}{{\left| {{A_3}} \right|}^2}{e_1}{e_3}} \right\}
\end{equation}
Since ${e_1} = 0$ and ${A_1} = A_1^* = 0$, $G_{(2)}^{13} = 0$,
\[G_{\left( 2 \right)}^{13} =  - \frac{1}{4}\frac{{P{m^{ - 1}}{H_2}\widetilde {Ra}}}{{{{\left| {{{\hat q}_2}} \right|}^2}{{\left| {{{\hat D}_{{W_2}}}} \right|}^2}{{\left| {{{\hat D}_{{H_2}}}} \right|}^2}}}\left\{ {\frac{{\hat q_2^*\hat D_{{W_2}}^*\hat D_{{\theta _2}}^* + {{\hat q}_2}{{\hat D}_{{W_2}}}{{\hat D}_{{\theta _2}}} + 2\widetilde {Ra}}}{{{{\left| {{{\hat q}_2}{{\hat D}_{{W_2}}}{{\hat D}_{{\theta _2}}} + \widetilde {Ra}} \right|}^2}}} - \frac{2}{{\widetilde {Ra}}}} \right\} + \]
\begin{equation}\label{e143}
   + \frac{i}{4}\frac{{\left( {1 - {W_2}} \right){H_2}\widetilde {Ra}}}{{{{\left| {{{\hat q}_2}} \right|}^2}{{\left| {{{\hat D}_{{W_2}}}} \right|}^2}{{\left| {{{\hat D}_{{H_2}}}} \right|}^2}}}\left\{ {\frac{{\hat q_2^*\hat D_{{W_2}}^*\hat D_{{\theta _2}}^* - {{\hat q}_2}{{\hat D}_{{W_2}}}{{\hat D}_{{\theta _2}}}}}{{{{\left| {{{\hat q}_2}{{\hat D}_{{W_2}}}{{\hat D}_{{\theta _2}}} + \widetilde {Ra}} \right|}^2}}}} \right\}
\end{equation}
\[G_{(1)}^{31} = \frac{{iP{m^{ - 1}}{H_1}}}{{{{\left| {{{\hat q}_1}} \right|}^2}{{\left| {{{\hat D}_{{W_1}}}} \right|}^2}{{\left| {{{\hat D}_{{H_1}}}} \right|}^2}}} \times\]
\[\times \left\{ {\left( {{A_1}A_3^* - {A_3}A_1^*} \right) - \hat aA_3^*\left( {{e_3}{A_1} - {e_1}{A_3}} \right) - {{\hat a}^*}{A_3}\left( {{e_1}A_3^* - {e_3}A_1^*} \right)} \right\} - \]
\[ - \frac{{(1 - {W_1}){H_1}}}{{{{\left| {{{\hat q}_1}} \right|}^2}{{\left| {{{\hat D}_{{W_1}}}} \right|}^2}{{\left| {{{\hat D}_{{H_1}}}} \right|}^2}}}\left\{ {\left( {{A_1}A_3^* + {A_3}A_1^*} \right) - \hat aA_3^*\left( {{e_3}{A_1} + {e_1}{A_3}} \right) - } \right.\]
\begin{equation}\label{e144}
  \left. {-  {{\hat a}^*}{A_3}\left( {{e_1}A_3^* + {e_3}A_1^*} \right) + 2{{\left| {\hat a} \right|}^2}{{\left| {{A_3}} \right|}^2}{e_1}{e_3}} \right\}
\end{equation}
because ${e_1} = 0$ and ${A_1} = A_1^* = 0$;
\[G_{\left( 2 \right)}^{31} = \frac{1}{4}\frac{{P{m^{ - 1}}{H_2}\widetilde {Ra}}}{{{{\left| {{{\hat q}_2}} \right|}^2}{{\left| {{{\hat D}_{{W_2}}}} \right|}^2}{{\left| {{{\hat D}_{{H_2}}}} \right|}^2}}}\left\{ {\frac{{\hat q_2^*\hat D_{{W_2}}^*\hat D_{{\theta _2}}^* + {{\hat q}_2}{{\hat D}_{{W_2}}}{{\hat D}_{{\theta _2}}} + 2\widetilde {Ra}}}{{{{\left| {{{\hat q}_2}{{\hat D}_{{W_2}}}{{\hat D}_{{\theta _2}}} + \widetilde {Ra}} \right|}^2}}} - \frac{2}{{\widetilde {Ra}}}} \right\} + \]
\begin{equation}\label{e145}
  + \frac{i}{4}\frac{{\left( {1 - {W_2}} \right){H_2}\widetilde {Ra}}}{{{{\left| {{{\hat q}_2}} \right|}^2}{{\left| {{{\hat D}_{{W_2}}}} \right|}^2}{{\left| {{{\hat D}_{{H_2}}}} \right|}^2}}}\left\{ {\frac{{\hat q_2^*\hat D_{{W_2}}^*\hat D_{{\theta _2}}^* - {{\hat q}_2}{{\hat D}_{{W_2}}}{{\hat D}_{{\theta _2}}}}}{{{{\left| {{{\hat q}_2}{{\hat D}_{{W_2}}}{{\hat D}_{{\theta _2}}} + \widetilde {Ra}} \right|}^2}}}} \right\}
\end{equation}
\[G_{(1)}^{23} = \frac{1}{4}\frac{{P{m^{ - 1}}{H_1}\widetilde {Ra}}}{{{{\left| {{{\hat q}_1}} \right|}^2}{{\left| {{{\hat D}_{{W_1}}}} \right|}^2}{{\left| {{{\hat D}_{{H_{123}}}}} \right|}^2}}}\left\{ {\frac{{\hat q_1^*\hat D_{{W_1}}^*\hat D_{{\theta _1}}^* + {{\hat q}_1}{{\hat D}_{{W_1}}}{{\hat D}_{{\theta _1}}} + 2\widetilde {Ra}}}{{{{\left| {{{\hat q}_1}{{\hat D}_{{W_1}}}{{\hat D}_{{\theta _1}}} + \widetilde {Ra}} \right|}^2}}} - \frac{2}{{\widetilde {Ra}}}} \right\} - \]
\begin{equation}\label{e146}
  - \;\frac{{\;\;i}}{4}\frac{{\left( {1 - {W_1}} \right){H_1}\widetilde {Ra}}}{{{{\left| {{{\hat q}_{12}}} \right|}^2}{{\left| {{{\hat D}_{{W_{12}}}}} \right|}^2}{{\left| {{{\hat D}_{{H_{12}}}}} \right|}^2}}}\left\{ {\frac{{\hat q_1^*\hat D_{{W_1}}^*\hat D_{{\theta _1}}^* - {{\hat q}_1}{{\hat D}_{{W_1}}}{{\hat D}_{{\theta _1}}}}}{{{{\left| {{{\hat q}_1}{{\hat D}_{{W_1}}}{{\hat D}_{{\theta _1}}} + \widetilde {Ra}} \right|}^2}}}} \right\}
\end{equation}
\[G_{(2)}^{23} = \frac{{iP{m^{ - 1}}{H_2}}}{{{{\left| {{{\hat q}_2}} \right|}^2}{{\left| {{{\hat D}_{{W_2}}}} \right|}^2}{{\left| {{{\hat D}_{{H_2}}}} \right|}^2}}} \times\]
\[\times \left\{ {\left( {{B_3}B_2^* - {B_2}B_3^*} \right) - \hat bB_3^*\left( {{e_2}{B_3} - {e_3}{B_2}} \right) - {{\hat b}^*}{B_3}\left( {{e_3}B_2^* - {e_2}B_3^*} \right)} \right\} - \]
\[ - \frac{{(1 - {W_2}){H_2}}}{{{{\left| {{{\hat q}_2}} \right|}^2}{{\left| {{{\hat D}_{{W_2}}}} \right|}^2}{{\left| {{{\hat D}_{{H_2}}}} \right|}^2}}}\left\{ {\left( {{B_3}B_2^* + {B_2}B_3^*} \right) - \hat bB_3^*\left( {{e_2}{B_3} + {e_3}{B_2}} \right) -  } \right.\]
\begin{equation}\label{e147}
  \left. { - {{\hat b}^*}{B_3}\left( {{e_3}B_2^* + {e_2}B_3^*} \right) + 2{{\left| {\hat a} \right|}^2}{{\left| {{B_3}} \right|}^2}{e_2}{e_3}} \right\} = 0
\end{equation}
because ${e_2} = 0$ and ${B_2} = B_2^* = 0$;
\[G_{\left( 1 \right)}^{32} =  - \frac{1}{4}\frac{{P{m^{ - 1}}{H_1}\widetilde {Ra}}}{{{{\left| {{{\hat q}_1}} \right|}^2}{{\left| {{{\hat D}_{{W_1}}}} \right|}^2}{{\left| {{{\hat D}_{{H_{123}}}}} \right|}^2}}}\left\{ {\frac{{\hat q_1^*\hat D_{{W_1}}^*\hat D_{{\theta _1}}^* + {{\hat q}_1}{{\hat D}_{{W_1}}}{{\hat D}_{{\theta _1}}} + 2\widetilde {Ra}}}{{{{\left| {{{\hat q}_1}{{\hat D}_{{W_1}}}{{\hat D}_{{\theta _1}}} + \widetilde {Ra}} \right|}^2}}} - \frac{2}{{\widetilde {Ra}}}} \right\} - \]
\begin{equation}\label{e148}
  - \frac{i}{4}\frac{{\left( {1 - {W_1}} \right){H_1}\widetilde {Ra}}}{{{{\left| {{{\hat q}_{12}}} \right|}^2}{{\left| {{{\hat D}_{{W_{12}}}}} \right|}^2}{{\left| {{{\hat D}_{{H_{12}}}}} \right|}^2}}}\left\{ {\frac{{\hat q_1^*\hat D_{{W_1}}^*\hat D_{{\theta _1}}^* - {{\hat q}_1}{{\hat D}_{{W_1}}}{{\hat D}_{{\theta _1}}}}}{{{{\left| {{{\hat q}_1}{{\hat D}_{{W_1}}}{{\hat D}_{{\theta _1}}} + \widetilde {Ra}} \right|}^2}}}} \right\}
\end{equation}
\[G_{(2)}^{32} = \frac{{iP{m^{ - 1}}{H_2}}}{{{{\left| {{{\hat q}_2}} \right|}^2}{{\left| {{{\hat D}_{{W_2}}}} \right|}^2}{{\left| {{{\hat D}_{{H_2}}}} \right|}^2}}}\times\]
\[\times \left\{ {\left( {{B_2}B_3^* - {B_3}B_2^*} \right) - \hat bB_3^*\left( {{e_3}{B_2} - {e_2}{B_3}} \right) - {{\hat b}^*}{B_3}\left( {{e_2}B_3^* - {e_3}B_2^*} \right)} \right\} - \]
\[ - \frac{{(1 - {W_2}){H_2}}}{{{{\left| {{{\hat q}_2}} \right|}^2}{{\left| {{{\hat D}_{{W_2}}}} \right|}^2}{{\left| {{{\hat D}_{{H_2}}}} \right|}^2}}}\left\{ {\left( {{B_2}B_3^* + {B_3}B_2^*} \right) - \hat bB_3^*\left( {{e_2}{B_3} + {e_3}{B_2}} \right) -  } \right.\]
\begin{equation}\label{e149}
  \left. {- {{\hat b}^*}{B_3}\left( {{e_3}B_2^* + {e_2}B_3^*} \right) + 2{{\left| {\hat a} \right|}^2}{{\left| {{B_3}} \right|}^2}{e_2}{e_3}} \right\}
\end{equation}
For the correlators components obtained here we  use the following relationships:
\[{\hat q_2}{\hat D_{{W_2}}}{\hat D_{{\theta _2}}} - \hat q_2^*\hat D_{{W_2}}^*\hat D_{{\theta _2}}^* = 2i\left( {1 + P{r^{ - 1}}} \right)\left( {1 - {W_2}} \right)+\]
\begin{equation}\label{e150}
   + \widetilde Q H_2^2 \left( \frac{{2i\left( {P{m^{ - 1}} - P{r^{ - 1}}} \right)(1 - {W_2})}}{{P{m^{ - 2}} + {{(1 - {W_2})}^2}}} \right)
\end{equation}
\[\hat q_2^*\hat D_{{W_2}}^*\hat D_{{\theta _2}}^* + {\hat q_2}{\hat D_{{W_2}}}{\hat D_{{\theta _2}}} = 2\left( {P{r^{ - 1}} - {{\left( {1 - {W_2}} \right)}^2}} \right) +\]
\begin{equation}\label{e151}
  + \frac{{2\widetilde QH_2^2(P{r^{ - 1}}P{m^{ - 1}} + {{\left( {1 - {W_2}} \right)}^2})}}{{P{m^{ - 2}} + {{(1 - {W_2})}^2}}}
\end{equation}
\[{\left| {{{\hat q}_2}{{\hat D}_{{W_2}}}{{\hat D}_{{\theta _2}}} + \widetilde {Ra}} \right|^2} =\]
\[ = {\left| {{{\hat q}_2}} \right|^2}{\left| {{{\hat D}_{{W_2}}}} \right|^2}{\left| {{{\hat D}_{{\theta _2}}}} \right|^2} + \widetilde {Ra}\left( {\hat q_2^*\hat D_{{W_2}}^*\hat D_{{\theta _2}}^* + {{\hat q}_2}{{\hat D}_{{W_2}}}{{\hat D}_{{\theta _2}}}} \right) + {\widetilde {Ra}^2} = \]
\[ = \left[ {\frac{{{{\left( {P{m^{ - 1}} - {{\left( {1 - {W_2}} \right)}^2} + \widetilde QH_2^2} \right)}^2} + {{\left( {1 - {W_2}} \right)}^2}{{\left( {1 + P{m^{ - 1}}} \right)}^2}}}{{\left( {1 + {{\left( {1 - {W_2}} \right)}^2}} \right)\left( {P{m^{ - 2}} + {{\left( {1 - {W_2}} \right)}^2}} \right)}}} \right]\times\]
\[\times \left( {1 + {{\left( {1 - {W_2}} \right)}^2}} \right)\left( {P{r^{ - 2}} + {{\left( {1 - {W_2}} \right)}^2}} \right) +  {\widetilde {Ra}^2}+\]
\begin{equation}\label{e152}
  + 2\widetilde {Ra}\left[ {\left( {P{r^{ - 1}} - {{\left( {1 - {W_2}} \right)}^2}} \right) + \frac{{\widetilde QH_2^2(P{r^{ - 1}}P{m^{ - 1}} + {{\left( {1 - {W_2}} \right)}^2})}}{{P{m^{ - 2}} + {{(1 - {W_2})}^2}}}} \right] ;
\end{equation}
Let us substitute equations (\ref{e150})-(\ref{e152}) in expressions for the components of $T_{(2)}^{31}$ and $T_{(1)}^{32}$. As a result we obtain:
\[T_{(2)}^{31} =  - \frac{{\widetilde {Ra}(1 + P{m^2}\widetilde W_2^2){{\widetilde W}_2}\left[ {\left(( 1 + \Pr )(1 + {P}{m^{2}}\widetilde {W}_{2}^{2}\;) \right) + QH_2^2(Pr - Pm)} \right]}}{{2\left[ {{{(1 - Pm\widetilde W_2^2 + QH_2^2)}^2} + \widetilde W_2^2{{(1 + Pm)}^2}} \right]}} \times \]
\[ \times \left[ {\left( {{{\left( {1 - Pm\widetilde W_2^2 + QH_2^2} \right)}^2} + \widetilde W_2^2{{\left( {1 + Pm} \right)}^2}} \right)\left( {1 + P{r^2}\widetilde W_2^2} \right) + } \right.\]
\[ + 2Ra\left( {\left( {1 - Pr\widetilde W_2^2} \right)\left( {1 + P{m^2}\widetilde W_2^2} \right) + QH_2^2\left( {1 + Pm\widetilde W_2^2} \right)} \right) + \]
\begin{equation}\label{e153}
  {\left. { + R{a^2}(1 + P{m^2}\widetilde W_2^2)} \right]^{ - 1}} =  - {\alpha ^{(2)}} \cdot {\widetilde W_2}
\end{equation}
\[T_{(1)}^{32} = \frac{{\widetilde {Ra}(1 + P{m^2}\widetilde W_1^2){{\widetilde W}_1}\left[ {\left( (1 + \Pr )(1 + {\rm{P}}{m^{\rm{2}}}\widetilde {\rm{W}}_{\rm{1}}^{2}\;) \right) + QH_1^2(Pr - Pm)} \right]}}{{2\left[ {{{(1 - Pm\widetilde W_1^2 + QH_1^2)}^2} + \widetilde W_1^2{{(1 + Pm)}^2}} \right]}} \times \]
\[ \times \left[ {\left( {{{\left( {1 - Pm\widetilde W_1^2 + QH_1^2} \right)}^2} + \widetilde W_1^2{{\left( {1 + Pm} \right)}^2}} \right)\left( {1 + P{r^2}\widetilde W_1^2} \right) + } \right.\]
\[ + 2Ra\left( {\left( {1 - Pr\widetilde W_1^2} \right)\left( {1 + P{m^2}\widetilde W_1^2} \right) + QH_1^2\left( {1 + Pm\widetilde W_1^2} \right)} \right) + \]
\begin{equation}\label{e154}
  {\left. { + R{a^2}(1 + P{m^2}\widetilde W_1^2)} \right]^{ - 1}} = {\alpha ^{(1)}} \cdot {\widetilde W_1}
\end{equation}
here ${\widetilde W_1} = 1 - {W_1}$, ${\widetilde W_2} = 1 - {W_2}$; ${\alpha ^{(1)}}$ and ${\alpha ^{(2)}}$ are coefficients of nonlinear hydrodynamic $\alpha$-effect in an electrically conductive medium with temperature stratification. Comparing the expressions (\ref{e133}) and (\ref{e139}), and (\ref{e135}) and (\ref{e140}), we find the connection of $S_{(2)}^{31}$ with $T_{(2)}^{31}$ and $S_{(1)}^{32}$ with $T_{(1)}^{32}$, i.e.
\begin{equation}\label{e155}
  S_{(2)}^{31} = \frac{{H_2^2T_{(2)}^{31}}}{{P{m^{ - 2}} + \widetilde W_2^2}};\quad S_{(1)}^{32} = \frac{{H_1^2T_{(1)}^{32}}}{{P{m^{ - 2}} + \widetilde W_1^2}}
\end{equation}
To close the equations of large-scale magnetic field (89), we need to calculate the turbulent e.m.f. $G_{(2)}^{13} - G_{(2)}^{31}$ and $G_{(1)}^{23} - G_{(1)}^{32}$. Taking into account the equations  (\ref{e143}), (\ref{e145}), (\ref{e146}), (\ref{e148}) we obtain:
\[\delta {G_{(2)}} = G_{(2)}^{13} - G_{(2)}^{31} =  - \frac{1}{2}\frac{{P{m^{ - 1}}{H_2}\widetilde {Ra}}}{{{{\left| {{{\hat q}_2}} \right|}^2}{{\left| {{{\hat D}_{{W_2}}}} \right|}^2}{{\left| {{{\hat D}_{{H_2}}}} \right|}^2}}} \times\]
\begin{equation}\label{e156}
  \times \left\{ {\frac{{\hat q_2^*\hat D_{{W_2}}^*\hat D_{{\theta _2}}^* + {{\hat q}_2}{{\hat D}_{{W_2}}}{{\hat D}_{{\theta _2}}} + 2\widetilde {Ra}}}{{{{\left| {{{\hat q}_2}{{\hat D}_{{W_2}}}{{\hat D}_{{\theta _2}}} + \widetilde {Ra}} \right|}^2}}} - \frac{2}{{\widetilde {Ra}}}} \right\}
\end{equation}
\[\delta {G_{(1)}} = G_{(1)}^{23} - G_{(1)}^{32} = \frac{1}{2}\frac{{P{m^{ - 1}}{H_1}\widetilde {Ra}}}{{{{\left| {{{\hat q}_1}} \right|}^2}{{\left| {{{\hat D}_{{W_1}}}} \right|}^2}{{\left| {{{\hat D}_{{H_1}}}} \right|}^2}}}\times\]
\begin{equation}\label{e157}
  \times\left\{ {\frac{{\hat q_1^*\hat D_{{W_1}}^*\hat D_{{\theta _1}}^* + {{\hat q}_1}{{\hat D}_{{W_1}}}{{\hat D}_{{\theta _1}}} + 2\widetilde {Ra}}}{{{{\left| {{{\hat q}_1}{{\hat D}_{{W_1}}}{{\hat D}_{{\theta _1}}} + \widetilde {Ra}} \right|}^2}}} - \frac{2}{{\widetilde {Ra}}}} \right\}
\end{equation}
After substituting the expressions (\ref{e151})-(\ref{e152}) in the equations (\ref{e156})-(\ref{e157}), we obtain the expression:
\[\delta {G_{(2)}} = \frac{{Pm{H_2}}}{{({{\left( {1 - Pm\widetilde W_2^2 + QH_2^2} \right)}^2} + \widetilde W_2^2{{\left( {1 + Pm} \right)}^2})}} \times \]
\[ \times \left\{ {1 - Ra\left[ {\left( {1 - Pr\widetilde W_2^2} \right) + \frac{{QH_2^2(1 + PrPm\widetilde W_2^2)}}{{(1 + P{m^2}\tilde W_2^2)}} + Ra} \right] \times } \right.\]
\[ \times \left[ {\left( {{{\left( {1 - Pm\widetilde W_2^2 + QH_2^2} \right)}^2} + \widetilde W_2^2{{\left( {1 + Pm} \right)}^2}} \right)\frac{{\left( {1 + P{r^2}\widetilde W_2^2} \right)}}{{\left( {1 + P{m^2}\widetilde W_2^2} \right)}} + } \right.\]
\[ \left. {{{\left. { + 2Ra\left[ {\left( {1 - Pr\widetilde W_2^2} \right) + \frac{{QH_2^2\left( {1 + PrPm\widetilde W_2^2} \right)}}{{\left( {1 + P{m^2}\widetilde W_2^2} \right)}}} \right] + R{a^2}} \right]}^{ - 1}}} \right\}=\]
\begin{equation}\label{e158}
  = \alpha _H^{(2)} \cdot {H_2};
\end{equation}
\[{G_{(1)}} = \frac{{ - Pm{H_1}}}{{({{\left( {1 - Pm\widetilde W_1^2 + QH_1^2} \right)}^2} + \widetilde W_1^2{{\left( {1 + Pm} \right)}^2})}}\left\{ {1 - \frac{}{}} \right.\]
\[ - Ra\left[ {\left( {1 - Pr\widetilde W_1^2} \right) + \frac{{QH_1^2(1 + PrPm\widetilde W_1^2)}}{{(1 + P{m^2}\widetilde W_1^2)}} + Ra} \right] \times \]
\[ \times \left[ {\left( {{{\left( {1 - Pm\widetilde W_1^2 + QH_1^2} \right)}^2} + \widetilde W_1^2{{\left( {1 + Pm} \right)}^2}} \right)\frac{{\left( {1 + P{r^2}\widetilde W_1^2} \right)}}{{\left( {1 + P{m^2}\widetilde W_1^2} \right)}} + } \right.\]
\[\left. {{{\left. { + 2Ra\left[ {\left( {1 - Pr\widetilde W_1^2} \right) + \frac{{QH_1^2\left( {1 + PrPm\widetilde W_1^2} \right)}}{{\left( {1 + P{m^2}\widetilde W_1^2} \right)}}} \right] + R{a^2}} \right]}^{ - 1}}} \right\} =\]
\begin{equation}\label{e159}
   =  - \alpha _H^{(1)} \cdot {H_1};
\end{equation}
Here $\alpha _H^{(1)}$, $\alpha _H^{(2)}$ are coefficients of nonlinear MHD $\alpha$-effect in an electrically conductive medium with temperature stratification. The coefficients of the nonlinear MHD $\alpha$-effect is responsible for the generation of large-scale magnetic fields and consist of two parts:
\begin{equation}\label{e160}
  \alpha _H^{(1)} = \alpha _H^{(0)}(1 - Ra \cdot \Phi \left( {{W_1},{H_1}} \right)),\;\;\alpha _H^{(2)} = \alpha _H^{(0)}(1 - Ra \cdot \Phi \left( {{W_2},{H_2}} \right))
\end{equation}
The first part of the $\alpha _H^{(0)}$ is determined only by the action of external helical force ${\vec f_0}$, the second part of the coefficients $\alpha _H^{(1)}$ and $\alpha _H^{(2)}$ are associated with the presence of the temperature stratification $Ra \ne 0$ if $\frac{{d{T_{00}}}}{{dz}} \ne 0$. Here $\Phi \left( {{W_{1,2}},{H_{1,2}}} \right)$ is  a certain function from ${W_{1,2}}$ and ${H_{1,2}}$.

\end{document}